\def\Granadadep{Departamento de F\'\i sica Te\'orica y del Cosmos, Facultad
de Ciencias, Universidad de Granada, Campus de Fuentenueva, Granada 18002,
Spain}

\def\Granadainst{Instituto de F\'\i sica Te\'orica y Computacional Carlos I,
Facultad
de Ciencias, Universidad de Granada, Campus de Fuentenueva, Granada 18002,
Spain}

\def\Valencia{IFIC, Centro Mixto Universidad de Valencia-CSIC, Burjasot
              46100-Valencia, Spain.}
\def\Comision{Work partially supported by the DGICYT.}

\def\nn{\nonumber}
\def\ni{\noindent}

\def\be{\begin{equation}}
\def\ee{\end{equation}}
\def\bea{\begin{eqnarray}}
\def\eea{\end{eqnarray}}
\def\ba{\begin{array}}
\def\ea{\end{array}}

\def\e{\hbox{{\large{\sl e}}}}
\def\z{\zeta}
\def\a{\alpha}
\def\b{\beta}
\def\av{\vec{\alpha}}
\def\bv{\vec{\beta}}
\def\ep{\epsilon}
\def\epv{\vec{\epsilon}}
\def\Gt{$\widetilde{G}\,\,$}
\def\Gtm{\widetilde{G}\,}

\def\GP{G_{\cal P}}
\def\GH{G_{\cal H}}
\def\GC{G_C}

\def\N{\hat{{\rm N}}(t)}
\def\Ni{\hat{{\rm N}}^{-1}(t)}
\def\M{\hat{{\rm M}}(t)}
\def\Mi{\hat{{\rm M}}^{-1}(t)}
\def\J{\hat{{\rm J}}}
\def\I{\hat{{\rm I}}}
\def\wc{\omega_c}
\def\Bmt{\wc t}
\def\Bm{\wc}
\def\r{\vec{r}}
\def\ro{\vec{r}_0}
\def\R{\vec{R}}
\def\xo{\vec{x}_0}
\def\n{\vec{n}}
\def\np{\vec{n}{}'}
\def\k{\vec{k}}
\def\kO{\vec{k}_0}
\def\kp{\vec{k}{}'}
\def\L{\vec{L}}

\def\Gthetam{{\cal G}_{\Theta}}

\def\x{\vec{x}}
\def\p{\vec{p}}
\def\z{\zeta}
\def\medio{\frac{1}{2}}
\def\E{\hat{{\rm E}}}
\def\P{\hat{{\rm P}}}
\def\Av{\hat{\vec{{\rm A}}}}
\def\Piv{\hat{\vec{\Pi}}}
\def\Tv{\hat{\vec{{\rm T}}}}
\def\Pv{\hat{\vec{{\rm P}}}}
\def\X{\hat{{\rm X}}}
\def\Xv{\hat{\vec{{\rm X}}}}

\def\en{{e_n}}

\def\ni{\noindent}
\def\idhbar{\frac{i}{\hbar}}
\def\w{\omega}

\newcommand{\parcial}[1]{ \frac{\partial}{\partial #1} }

\newcommand{\XL}[1]{ {\tilde{X}}^{L}_{#1} }

\newcommand{\XR}[1]{ {\tilde{X}}^{R}_{#1} }

\newcommand{\iXR}[1]{ i_{\XR{#1}}\Theta}

\documentstyle[12pt,epsf]{article}

\begin{document}



\begin{center}
{\Large {\bf ALGEBRAIC QUANTIZATION, GOOD OPERATORS AND FRACTIONAL
      QUANTUM NUMBERS$^1$ }}
\end{center}

\bigskip
\bigskip

\centerline{ V. Aldaya$^{2,3}$, M. Calixto$^2$
              and J. Guerrero$^{2,4}$  }       

\bigskip
\centerline{September 20, 1995}
\bigskip

\footnotetext[1]{\Comision}
\footnotetext[2]{\Granadainst} \footnotetext[3]{\Valencia}
\footnotetext[4]{\Granadadep}

\bigskip

\begin{center}
{\bf Abstract}
\end{center}

\small

\begin{list}{}{\setlength{\leftmargin}{3pc}\setlength{\rightmargin}{3pc}}
\item The problems arising when quantizing systems with periodic boundary
conditions are analysed, in an algebraic (group-) quantization scheme,
and the ``failure" of the Ehrenfest theorem is clarified in terms of the
already defined notion of {\it good} (and {\it bad}) operators. The
analysis of ``constrained" Heisenberg-Weyl groups according to this
quantization scheme reveals the possibility for new quantum (fractional)
numbers extending those allowed for Chern classes in traditional
Geometric Quantization. This study is illustrated with the examples
of the free particle on the circumference and the charged particle in a
homogeneous magnetic field on the torus, both examples featuring
``anomalous" operators, non-equivalent quantization and the latter,
fractional quantum numbers. These provide the rationale behind flux
quantization in superconducting rings and Fractional Quantum Hall
Effect, respectively.
\end{list}

\normalsize


\section{Introduction}

The need for a consistent quantization scheme which is truly suitable for
systems
wearing a non-trivial topology is increasing daily. Configuration
spaces with non-trivial topology appear in as diverse cases as Gauge
Theories, Quantum Gravity, and the more palpable ones of the superconducting
ring and the Quantum Hall effect, where the measuring tools change
the topology of the system in a non-trivial way.

The most common problem which appears when the configuration-space
manifold possesses a non-trivial topology is the failure of the Ehrenfest
theorem for certain operators, a problem usually referred to as an anomaly.
In the sequel, we shall add the qualifier {\it topologic} to distinguish
these from others directly attached to the Lie algebra of the quantum
operators and characterized, roughly speaking, by the appearance of
a term in a quantum commutator not present at the classical,
Poisson-algebra level. We call them {\it algebraic} anomalies and refer
the reader to \cite{Schrodinger} for a detailed analysis.

The failure of the Ehrenfest theorem for a given operator is primarily
related to the non-globality of the corresponding classical function,
such as is the case of the local co-ordinate on a one-dimensional closed
submanifold.
Geometric Quantization was intended to go further than ordinary canonical
quantization does, allowing for the quantization of arbitrary symplectic
manifolds.  Unfortunately, Geometric Quantization only partially accomplished
this task, one of the reasons being the difficulty in
(or, even more, the impossibility of) finding a polarization suitable
enough to quantize a given set of classical functions \cite{Woodhouse,Isham},
or quantizing a set of operators in a way that would preserve a given
polarization.

A quantization procedure based on a group structure, Group Approach
to Quantization (GAQ) \cite{GAQ,Schrodinger}, improves
the standard Geometric Quantization approach in that it provides
two sets of mutually commuting operators, namely, the left- and
right-invariant vector fields.
This enables us to impose the polarization conditions by means of the
left-invariant vector fields, say, while the right-invariant ones will
be the quantum operators, which automatically preserve the polarization.
The quantization group \Gt is endowed with a $U(1)$-principal bundle structure
so that generators fall into two classes according to whether or not they
give rise to a term proportional to the vertical generator on the r.h.s. of
a commutator.  Generators which do not reproduce any $U(1)$-term close
a {\it horizontal}\  subalgebra, the characteristic subalgebra, of
non-dynamical generators, which should be included in the polarization
subalgebra.
The principal drawback of GAQ is the need for a group symmetry
associated with
the system to be quantized, and the apparent restriction in the number of
functions which can be quantized. However, this
last limitation
is slighter than it might seem, since Canonical Quantization on a particular
phase space does not quantize the entire set of functions on phase space, but
rather, a restricted Poisson subalgebra. Even
more, it could well happen in some cases that a more standard quantization
provides only quantum operators corresponding to a finite-dimensional Lie
algebra. This is the case, for instance, of the symplectic manifold $S^2$,
where the quantum operators are only those of $su(2)+R$
\cite{Gotay}. Moreover, most of the interesting systems in Physics possess a
symmetry group
large enough to achieve a proper quantization.

To be more precise, not only the right invariant vector fields preserve the
polarization, but rather the entire right enveloping algebra preserve the
structure of the Hilbert space. This means that any element in the right
enveloping algebra can be realized as a quantum operator, although the
relation between the quantum algebra and the standard Poisson algebra on the
co-adjoint orbits of the group is no longer an isomorphism; GQA provides
a quantum theory rather than the quantization of a classical theory.

A reformulation of GAQ was proposed a few years ago \cite{Alfonso}, the
Algebraic Quantization on a Group (AQG) [some of the basic ideas in
\cite{Alfonso} have also appeared in the context of quantum systems with
non-trivial topology \cite{Landsman} and in Quantum Gravity (\cite{Ashtekar}
and references therein)]
, which generalizes GAQ in
two respects. Firstly, finite transformations generalize
the infinitesimal ones throughout the method; that is, any concept or condition
relative to Lie subalgebras is generalized by its counterpart in terms of Lie
subgroups,
thus allowing discrete transformations to enter the theory. Needless to say,
infinitesimal objects are employed whenever possible. Secondly,
it generalizes the $U(1)$ phase invariance in Quantum
Mechanics (the structure group of the principal bundle fibration of the
quantum symmetry) incorporating other symmetries, eventually interpreted
as constraints. The new structure group $T$, which must include the
traditional $U(1)$, may also contain discrete symmetries especially
suitable to simulate manifold surgery as, for instance, toral
compactification, by means of periodic boundary conditions.

{}From now on we shall call ``compactified" (cylindrical or toral)
Heisenberg-Weyl (H-W) group a H-W group where the structure group is $T$
rather than $U(1)$, this subgroup  $T$ being the factor subgroup
leading to a compactified classical (the cylinder or the torus) phase space
by the quotient $\Gtm/T$.

The generalization of the $U(1)$-equivariance to $T$-equivariance
condition on the wave functions gives rise to two new, closely related
features: a) the existence of non-equivalent quantizations associated
with non-equivalent representations of the larger structural subgroup
$T$, and b) the notion of {\it good} operators, constituting the
subgroup of transformations compatible with the $T$-equivariance
condition, in a sense to be specified later (see \cite{Alfonso}).
Furthermore, those operators not preserving the $T$-equivariance
condition, the {\it bad} operators, may be seen as
quantization-changing transformations, and exhibit topologic
anomalies. Like in the $T=U(1)$ case, all the elements of the right enveloping
algebra {\it compatible with the $T$-equivariance condition}, for arbitrary
$T$, can be realized as {\it good} quantum operators.

It should be noted that, as mentioned above, AGQ is formulated in terms
of finite objects. This means that some algebraic indices must replace
the well-know Chern class $[\omega]$ of the symplectic form in Geometric
Quantization. In fact, the indices characterizing the
(not necessarily central) extension by $T$ of the ``classical" group $G$
generalize the Chern class, providing also {\it fractional} values. This
is precisely the case of the motion of a charged particle on a torus
in the presence of a homogeneous magnetic field, closely related to the
(Fractional) Quantum Hall Effect. The appearance of fractional quantum numbers
generalizing the integer Chern classes reveals, once again, that the
procedure of taking constraints and that of quantizing, depending at least
on the specific methods employed, may not commute.

This paper is organized as follows. Section 2 illustrates the way in which AQG
operates with the help of the examples of the Heisenberg-Weyl group in 1D with
constraints mimicking the compactification of the coordinate $x$  (Sec 2.1) and
that of the compactification of both $x$ and $p$ (Sec. 2.2). In the latter
case,
generalizing the quantization of a compact phase space {\it \`a la Dirac}, a
not necessarily integer quantization condition is obtained which generalizes
that of Geometric Quantization, i.e. the condition $[\omega]\in Z$
(Chern class), and, associated with it, vector-valued wave functions. In
solving this problem, real (versus holomorphic) polarizations
have been employed, leading to a generalized $kq$-representation. This
technique simplifies the treatment and is much more
intuitive, even though the configuration-space wave functions contain delta
functions. The results obtained in Sec. 2 are applied to the
quantization of the free particle on the circumference (Sec. 3, where the
failure of Ehrenfest theorem is analysed), directly related to flux
quantization in superconducting rings, and to the
quantization of a charged particle on a torus in the presence of an homogeneous
transverse magnetic field (Sec. 4), providing the rationale behind
Integer and Fractional Quantum Hall Effect.

\section{Algebraic Quantization of ``compactified" Heisenberg -Weyl
groups: Fractional quantum numbers}

In this section, we shall explain the AQG formalism over the
example of the Heisenberg-Weyl group with one co-ordinate ``compactified", i.e.
with constraints associated with the compactification of one co-ordinate,
and with one co-ordinate and its canonically conjugate momentum
``compactified". We nevertheless recommend the reading of the Ref.
\cite{Alfonso}. Although explicit calculations are given for the
Heisenberg-Weyl group with only one co-ordinate-momentum pair,
the results can be generalized, immediately, to any finite
number of them.

\subsection{Cylindrical Heisenberg-Weyl group}

Let us firstly proceed with the case of the Heisenberg-Weyl group with
only one of the coordinates ``compactified", i.e. with structure group $T$ such
that the quotient $\Gtm/T$ leads to the cylinder
as the symplectic manifold. The starting point in AQG is a Lie group
\Gt which is a right-principal
bundle with structure group $T$. $T$ is itself a principal bundle
with $U(1)$ as structure group. In our case  \Gt is
the ordinary Heisenberg-Weyl group in 1D (throughout the
paper, 1D means one coordinate-momentum pair, $x$ and $p$), and
$T=U(1)\times\{e_k,\,k\in Z\}$, where $\{e_k,\,k\in Z\}$ is the
subgroup of \Gt of finite translations in the coordinate $x$ by an
amount of $kL$, $L$ being
the spatial period. Note that $T$ is isomorphic to $U(1)\times Z$, so that
its fibration is trivial.

The group law $g''=g'*g$ for \Gt is:
\bea
x'' & = &   x'+ x \nn \\
p'' & = & p' + p   \label{H-WcGLaw} \\
\z''   & = &   \z'\z \e^{\frac{i}{\hbar}[(1+\lambda)x'p+\lambda xp']} \nn
\eea

\ni where the first two lines correspond to the group law of $G$, and
the third to that of $U(1)$. The real parameter $\lambda$ has
been introduced to account for a complete class of central extensions
differing in a coboundary [coboundaries have the form
$\xi(g',g)=\eta(g'*g)-\eta(g')-\eta(g)$, where $\eta:G\rightarrow R$ is called
the generating function of the coboundary]
generated by the function $\eta(x,p)=\lambda xp$
(In particular, for $\lambda=-\medio$ we have Bargmann's cocycle).

{}From this group law we can read immediately the right and left translations,
$R_{g'}g=g'*g=L_gg'$. In particular, the left- and right-invariant vector
fields (generating the finite translations) become:
\be
\ba{lcl}
\XL{x} & = & \parcial{x} + \frac{\lambda}{\hbar}p\,\Xi  \\
\XL{p}& = & \parcial{p} + \frac{1+\lambda}{\hbar}x\,\Xi \\
\XL{\z} & = & i\z\parcial{\z} \equiv \Xi
\ea \,\,\,\,\,\,
\ba{lcl}
\XR{x}& = & \parcial{x} + \frac{1+\lambda}{\hbar}p\,\Xi \\
\XR{p}& = & \parcial{p} + \frac{\lambda}{\hbar}x\,\Xi  \\
\XR{\z} & = & i\z\parcial{\z} \equiv \Xi
\ea \label{XLRH-Wc}
\ee

\ni The quantization 1-form (the left-invariant 1-form associated with the
parameter $\z$) can also be obtained:
\be
\Theta = - \lambda pdx - (1+\lambda)xdp + \hbar\frac{d\z}{i\z}
\label{1formC}
\ee

Since we are not considering time evolution, the quantization 1-form
has no characteristic subalgebra, there exists no discrete characteristic
subgroup $\GC$, and any combination of the two
generators $\XL{x}$ and $\XL{p}$ constitutes a first-order full
polarization (with the time evolution added, as in the free particle in
1D, things are a bit more complicated, see Sec. 3). Two polarizations
are singled out, ${\cal P}_p = <\XL{x}>$ and ${\cal P}_x = <\XL{p}>$,
or their finite (versus infinitesimal) counterparts
$\GP{}_p=\{\hbox{Space translations}\}$ and
$\GP{}_x=\{\hbox{Boosts transformations}\}$, leading to momentum and
configuration space representations, respectively. It should be borne in mind
that
the polarization conditions are needed to reduce the group representation
which otherwise would provide only the Bohr-Sommerfeld quantization.
These polarization conditions
read, in general, $\XL{}\Psi=0,\,\forall \XL{}\in {\cal P}$ or
$\Psi(g*\GP)=\Psi(g)$ in finite terms.

The $T$-function condition generalizes ordinary phase invariance
($U(1)$-equivariance) in Quantum Mechanics, which is written
$\Psi(\z*g)=\rho(\z)\Psi(g)$, where $\rho(\z)$ is the natural representation
of $U(1)$ on the complex numbers, $\rho(\z)=\z$. The generalization to a bigger
group $T$ involves the use of a general representation ${\cal D}$ of $T$ (or,
to be precise, of $T_B\equiv U(1)\cup T_p$, where $T_p$ is a maximal
polarization
subgroup of $T$; see \cite{Alfonso}) on a complex vector space $E$, where the
wave functions themselves take their values. In the formalism of AQG,
the representation of $T_B$ is constructed from the very representation of \Gt,
i.e. the vector space $E$ on which the constrained functions are evaluated is
made out of the unconstrained wave functions by properly choosing their
arguments.
This is the reason why the group $T_B$ is interpreted as constraints: the
representation of $T_B$ is not an abstract representation, but rather built
with
the same functions of the representation of \Gt.

The $T_B$-function condition then reads
$\Psi(g_{T_B}*g)={\cal D}(g_{T_B})\Psi(g), \forall g_{T_B}\in T_B, \forall
g\in \Gtm$. In the present
case $T_B=T$ and ${\cal D}(\z,e_k)=\z D(e_k)$, where $D(e_k)$ is a
representation of
$\{e_k,\,k\in Z\}$ ($\approx Z$) in the complex numbers, and
there is an infinity of non-equivalent irreducible representations, of the
form $D^\ep(e_k)=\e^{\idhbar \ep kL}$, with $\ep\in [0,\frac{2\pi\hbar}{L})$
(the first
Brillouin zone, in Solid State nomenclature). Therefore, {\it there is a
non-equivalent quantization associated with each choice of non-equivalent
representation of $T$, parameterized by $\ep$}. The $T$-function
condition for the wave function implies the restriction:
\be
\e^{\idhbar (1+\lambda)kLp}\Psi^\ep(x+kL,p,\z)=\e^{\idhbar \ep kL}
\Psi^\ep(x,p,\z) \label{H-WcConstraint}
\ee

\ni Note that the constrained wave functions can be identified with the space
of sections of a $U(1)$-bundle on the cylinder, with connection given by
(\ref{1formC}).

We now impose the polarization conditions in order to reduce the
representation.
Firstly, we shall consider the momentum space representation, where
the polarization conditions (either in finite or infinitesimal form)
lead to the following form of the wave functions:
\be
\Psi^\ep(x,p,\z) = \z \e^{-\idhbar \lambda xp} \Phi^\ep(p)
\label{WFH-WcMom}
\ee

\ni where the fact that $\Psi(\z g)=\z\Psi(g)$ (by the $T$-function
property) has been used.

\ni Both conditions (\ref{H-WcConstraint}) and (\ref{WFH-WcMom}) together
imply for the wave function $\Phi^\ep(p)$ a form like:
\be
\Phi^\ep(p)=\sum_{k\in Z} \a_k \phi^\ep_k(p)
\label{WFH-WcMom2}
\ee

\ni where $\phi^\ep_k(p)\equiv\delta(p-\ep-\frac{2\pi\hbar}{L}k)$,
i.e. the wave function is peaked at the values of the momentum
$p_k^\ep=\ep+\frac{2\pi\hbar}{L}k$, $k\in Z$. The Hilbert space ${\cal
H}^\ep(\Gtm)$
is made from the wave functions defined by (\ref{WFH-WcMom}) and
(\ref{WFH-WcMom2}).

The quantum operators, defined as $\P\equiv -i\hbar\XR{x}\,\,$ and
$\X\equiv i\hbar\XR{p}$, act on the wave functions as:
\bea
\P \Psi^\ep & = & p \Psi^\ep   \nn   \\
\X \Psi^\ep & = & \z \e^{-\idhbar \lambda xp}
           \left[i\hbar \parcial{p}\right] \Phi^\ep \label{H-WcOMom}
\eea

One of the main consequences of having generalized the structure group in
AQG is the classification of the operators (actually left translations) as
good and bad operators according to whether or not they are compatible with
the $T$-function condition. More precisely, the subgroup of good operators,
$\GH$, is characterized by the condition (see \cite{Alfonso})
\be
Ad(\Gtm)\left[g_T,\GH\right]\subset\GP\,,\;\; \forall g_T\in T
\label{Good}
\ee

In the present case, and
due to the discrete character of the ``physical" momenta, the position
operator $\X$ is expected to be problematic, since
the subgroup of good transformations
compatible with (\ref{H-WcConstraint}) and (\ref{WFH-WcMom}) is the
subgroup of \Gt, in which the continuous variable $p$ is  substituted by the
discrete variable $p_k\equiv p_k^0=\frac{2\pi\hbar}{L}k$, $k\in Z$, as
can be deduced
from $Ad(\Gtm)\left[\en,g\right]=(0,0,\e^{\idhbar nLp})\subset\GP\,\,
\forall n\in Z$. Therefore, the good operators are $\P$ and the finite boosts
transformations by the amount of $p_k$. Position  is not a
good operator in the sense that it does not preserve the structure
of the wave functions, i.e. it does not leave the Hilbert space
(for fixed $\ep$) ${\cal H}^\ep(\Gtm)$ stable. This fact will be further
discussed in Sec. 2.1.1.

With regard to the configuration space representation given by the
polarization ${\cal P}_x$ or the polarization subgroup $\GP{}_x$,
the solutions to this polarization are:
\be
\Psi(x,p,\z) = \z \e^{-\idhbar(1+\lambda) px} \Phi(x) \label{WFH-WcConf}
\ee

\ni Applying the condition
of $T$-function (\ref{H-WcConstraint}) to this wave functions in configuration
space,  we obtain:

\be
 \e^{\idhbar (1+\lambda)kLp} \e^{-\idhbar(1+\lambda)(x+kL)p}\Phi^\ep(x+kL) =
   \e^{\idhbar \ep kL}\e^{-\idhbar (1+\lambda)xp}\Phi^\ep(x)
\ee

\ni $\forall k\in Z$, where the quasi-periodicity condition for $\Phi^\ep(x)$
immediately follows:
\be
\Phi^\ep(x+L)=\e^{\idhbar \ep L}\Phi^\ep(x) \label{H-WcPeriod}
\ee

\ni It should be stressed that this result is independent of the chosen
cocycle, since it does not depend on $\lambda$, as expected.

The quantum operators are:
\bea
\P \Psi^\ep & = & \z \e^{-\idhbar(1+\lambda) px} \left[-i\hbar\nabla
       \right]\Phi^\ep \nn   \\
\X \Psi^\ep & = & \z x \Psi^\ep \label{H-WcOConf}
\eea

\ni Again, the position operator $\X$ is not a good operator, for the
same reason as in the momentum-space case, and the subgroup of (left)
transformations leaving the structure of the wave functions (\ref{WFH-WcConf})
and (\ref{H-WcPeriod}) stable is the same $\GH$ as
before, containing only $\P$ and the finite boosts in $p_k$, $k\in Z$.
Therefore,
the standard position has no meaning for any (Galilean) system with the
circumference as configuration space (see Sec. 3).

\subsubsection{Is there any good position-like operator?}

The position operator $\X$ is not a good operator because the variable
$x$ is not periodic: if $\phi(x)$ is a quasi-periodic function, $x\phi(x)$
is no longer quasi-periodic. However, the function $\eta=\e^{i\frac{2\pi}{L}x}$
is periodic, so that we could define the operator
$\hat{\eta}\equiv\e^{i\frac{2\pi}{L}\X}$, and verify that
$\hat{\eta}\Psi^\ep=\e^{i\frac{2\pi}{L}x}\Psi^\ep$ satisfies the same
quasi-periodicity condition as $\Psi^\ep$. We can then say that
$\hat{\eta}$ is a good operator.

The reason why $\hat{\eta}$ is a good operator is precisely that
it generates a good finite boost. We know that
the only good boosts are indexed by $p_k=\frac{2\pi\hbar}{L}k$, i.e.
\be
\Psi^\ep(p_k*g)=\e^{\frac{2\pi\hbar}{L}k\XR{p}}\Psi^\ep(g)=
\left(\e^{i\frac{2\pi}{L}\X}\right)^{-k}\Psi^\ep(g)=\hat{\eta}^{-k}\Psi^\ep(g)
\ee

\ni This means that $\hat{\eta}^k,\,k\in Z$ are the only good position
operators.

The finite operator $\hat{\eta}$ is obviously not Hermitian; rather, it is
unitary as
it should be. However $\hat{\eta}$ can be written as
$\hat{\eta}=\cos(\frac{2\pi}{L}\X) + i\sin(\frac{2\pi}{L}\X)$, the good
operators $\cos(\frac{2\pi}{L}\X)$ and $\sin(\frac{2\pi}{L}\X)$ being
Hermitian. These are good operators, given that they are
periodic functions of the operator $\X$. Since the set of functions
$\{\e^{i\frac{2\pi}{L}mx},\,m\in Z\}$ constitutes a basis for the periodic
functions of $x$ in the interval $[0,L]$, any operator which is a periodic
function of the position operator $\X$ is a good operator.

In any case, we might wonder about the finite boosts transformations
for $\tilde{p}\neq p_k$, i.e. about transformations of the form
$\Phi'(x)=\e^{\idhbar \tilde{p}\X}\Phi^\ep(x)=\e^{\idhbar \tilde{p}x}
\Phi^\ep(x)$. This new function verifies the boundary conditions
$\Phi'(x+L)=\e^{\idhbar \tilde{p}(x+L)}\Phi^\ep(x+L)=
 \e^{\idhbar (\ep+\tilde{p})L}\Phi'(x)$, and therefore  belongs to the
Hilbert
space ${\cal H}^{\ep+\tilde{p}}(\Gtm)$. In fact, since the representations
parameterized
by $\ep$ and $\ep+\frac{2\pi\hbar}{L}k, \,k\in Z$ are equivalent, the
transformed wave functions lie in the representation
$(\ep+\tilde{p}) \hbox{ mod } \frac{2\pi\hbar}{L}$. Of course, if
$\tilde{p}=p_k$ for some $k$,
the transformed wave function lies in the same Hilbert space as before and
we recover the result that the finite boosts in $p_k$ are good operators.
In particular,
$\Phi^\ep(x)=\e^{\idhbar \ep\X}\Phi^0(x)=\e^{\idhbar \ep x}\Phi^0(x)
\equiv \e^{\idhbar \ep x}\Phi(x)$, with $\Phi(x)$ satisfying (the usual)
periodic boundary conditions. This means that all the Hilbert spaces
${\cal H}^{\ep}(\Gtm)$, although yielding non-equivalent representations,
are related to each other by means of finite boosts transformations, which
are unitary transformations considered in the union of all these Hilbert
spaces $\cup_{\ep\in[0,2\pi\hbar/L)}{\cal H}^{\ep}(\Gtm)$.
We could say that
${\cal H}^{\ep}(\Gtm)$ for a fixed $\ep$ is too small for the boosts operator
to live in. The momentum operator, however, preserves (and is Hermitian in)
each one of these Hilbert spaces, but it is not Hermitian in the union of all
of them.

It is worth mentioning that the set of operators $\P$, $\hat{\eta}$ and
$\hat{\eta}^{\dag}$ close a  Lie algebra under ordinary commutation which is
isomorphic to the non-extended harmonic oscillator algebra. The operators
$\hat{\eta}$ and $\hat{\eta}^{\dag}$ act as ladder operators on the
eigenfunctions of $\P$ (this fact has been used in \cite{O-K} to study
Quantum Mechanics on the circumference).

\subsection{Toral Heisenberg-Weyl group}

Let us now proceed with the case of the Heisenberg-Weyl group with
both the coordinate and the momentum ``compactified", i.e. with a structure
group $T$ such that $\Gtm/T$ leads to the torus as the
symplectic manifold. We shall parameterize the plane with coordinates
$(x_1,x_2)$ because in physical applications the coordinates play the double
r\^ole of coordinate and momentum (see Sec. 4.1).

We also apply AQG to this system. Here again, \Gt is
the Heisenberg-Weyl group in 1D (now parameterized by $\x=(x_1,x_2)$ and $\z$).
Given $L_1$ and $L_2$, we introduce the lattice points
$\L_{\k}\equiv (k_1 L_1,k_2 L_2)$, $k_1$ and $k_2$ being integers
(thus defining a rectangular torus if we would take quotient by them).
The structure group $T$ will be a principal bundle with base
$\left\{e_{\k},\,\k\in Z\times Z \right\}$ and fiber $U(1)$, where
$\left\{e_{\k},\,\k\in Z\times Z \right\}\subset \Gtm$ is the
set of finite translations in the coordinates $\x$ by
an amount of $\L_{\k}$. The fibration of $T$ is non-trivial in general.

The group law for \Gt now reads:
\bea
\x{\:}'' & = &   \x{\:}'+ \x \nn \\
\z''   & = &   \z'\z \e^{\idhbar m\w[(1+\lambda)x_1'x_2+\lambda x_1x_2']}
 \label{H-WtGLaw}
\eea

\ni where a new numerical constant $\w$ (with dimensions of $T^{-1}$),
besides the mass $m$, which was implicit in the momentum $p=mv$, has been
introduced to accommodate the dimensions in the exponent above.

The left and right invariant vector fields can be obtained:
\be
\ba{lcl}
\XL{x_1} & = & \parcial{x_1} + \frac{\lambda}{\hbar}m\w x_2\,\Xi  \\
\XL{x_2}& = & \parcial{x_2} + \frac{1+\lambda}{\hbar}m\w x_1\,\Xi \\
\XL{\z} & = & i\z\parcial{\z} \equiv \Xi
\ea \,\,\,\,\,\,
\ba{lcl}
\XR{x_1}& = & \parcial{x_1} + \frac{1+\lambda}{\hbar}m\w x_2\,\Xi \\
\XR{x_2}& = & \parcial{x_2} + \frac{\lambda}{\hbar}m\w x_1\,\Xi  \\
\XR{\z} & = & i\z\parcial{\z} \equiv \Xi \,,
\ea \label{XLRH-Wt}
\ee

\ni and the quantization 1-form is:
\be
\Theta = - \lambda m\w x_2dx_1 - (1+\lambda)m\w x_1dx_2 + \hbar\frac{d\z}{i\z}
\label{1formT}
\ee

\ni As before, the quantization 1-form
has no characteristic module, and any combination of the two
generators $\XL{x_1}$ and $\XL{x_2}$ constitutes a first-order full
polarization. These can be written as ${\cal P}_{\n} = < \n\cdot\XL{\x}>$,
where $\n=(n_1,n_2)$ is an arbitrary unit vector. The choice of an $\n$
corresponds to the selection of a particular direction in the plane.
[All directions are indistinguishable, but on the mimicked torus, there are
geodesics (directions) which close, as happens with
the lines $x_2=0$ and $x_1=0$, and others which are open and fill
the torus densely. It can be easily checked that the condition for a geodesic
with
direction given by $\n$ to close is either that
$\frac{n_2}{n_1}=\frac{k_{02}}{k_{01}} \frac{L_2}{L_1}\,,k_{01},k_{02}\in Z$
or that $\n=(1,0)$ or $\n=(0,1)$, i.e. $\n$ is of the form
$\n=\L_{\kO}/|\L_{\kO}|$, with $\kO\in Z\times Z$. Also,  for
a geodesic and its orthogonal one to close, it is necessary and sufficient that
$\frac{L_2^2}{L_1^2}$ be a rational, except for the case $\n=(1,0)$ and
$\n=(0,1)$, which are always orthogonal and closed. This
condition is similar to the condition of commensurability of the frequencies
for a Lissajoux figure to be closed].

The polarization condition ${\cal P}_{\n}$ leads to the following wave
functions:
\be
\Psi=\z \e^{-\idhbar m\w[ (\lambda n_1^2-(1+\lambda)n_2^2)y_1y_2 +
     (\lambda + \medio)n_1n_2 y_1^2]} \Phi(y_2) \label{H-WtWF}
\ee

\ni where $y_1\equiv \n\cdot\x\,,\,\, y_2\equiv \n\cdot\J\cdot\x$, and
$(\J)_{ij}=\epsilon_{ij}\,$, with $\epsilon_{12}=1$. The action of the right
operators  on these wave functions is:
\bea
\n\cdot\XR{\x}\Psi &=& \idhbar m\w y_2 \Psi \nn \\
\n\cdot\J\cdot\XR{\x}\Psi &=& \z \e^{-\idhbar m\w[ (\lambda
n_1^2-(1+\lambda)n_2^2)y_1y_2 +
     (\lambda + \medio)n_1n_2 y_1^2]}\times \nn \\
  & &  \left[\parcial{y_2} -\idhbar m\w n_1n_2(1+2\lambda)y_2\right]\Phi(y_2)
\label{H-WtOp}
\eea

Before imposing the constraints, we have to determine the structure
of the group $T$. It must be done by means of finite transformations,
since it is basically a
discrete group (times $U(1)$). We then compute the group commutator of two
elements of $\left\{e_{\k}\right\}_{\k\in Z\times Z}$, with the result
$[e_{\kp},e_{\k}]=(0,0,\e^{\idhbar m\w L_1L_2(k_1'k_2-k_2'k_1)})$. Two cases
have to be considered:
\begin{itemize}
 \item[{\it i)}] \,\, $\e^{\idhbar m\w L_1L_2(k_1'k_2-k_2'k_1)}=1 \,\forall\,
  \k,\,\kp \in Z\times Z \;\Rightarrow \; [e_{\kp},e_{\k}]={\bf1}_{\Gtm}$\, ,
\item[{\it ii)}] \,\,$\exists\,\k$ and $\kp$\, /\,
       $\e^{\idhbar m\w L_1L_2(k_1'k_2-k_2'k_1)}\neq 1\; \Rightarrow \;
     {\bf 1}_{\Gtm}\neq [e_{\kp},e_{\k}]\in U(1)$\, ,
\end{itemize}

For the case {\it i)}, $T$ is the direct product
$T=\left\{e_{\k},\,\k\in Z\times Z\right\}\times U(1)$ and the whole group
$T$ can be imposed as constraints.
For the case {\it ii)}, when $\e^{\idhbar m\w L_1L_2(k_1'k_2-k_2'k_1)}\neq 1$
for some values of $\k$ and $\kp$ (an infinite discrete set of values, in
fact), there are two possibilities, depending on whether
$\frac{m\w L_1L_2}{2\pi\hbar}$ is rational or irrational. In neither
case can we impose the entire group $T$ as a
constraint group and we have to choose a polarization subgroup $T_p$ of $T$
(see \cite{Alfonso}).

\subsubsection{Integer quantum numbers}

For the condition {\it i)} to hold, it is necessary that
\be
\frac{m\w L_1L_2}{2\pi\hbar}=n\in N \label{H-WtQuant} \, ,
\ee

\ni which implies a {\it quantization} of the ``frequency" $\w$. As we shall
see in Sec. 4, this condition will imply the quantization of the magnetic flux
through the torus surface. This quantization condition is of the same nature as
that of the Dirac monopole case. Concerning this case, AQG simply
reproduces the quantization condition of the standard Geometric Quantization:
the symplectic form must be of integer class, defining the Chern class of the
quantum manifold.

The rest of the procedure follows the
same lines as in the case of the cylindrical H-W group: the condition
of $T$-function is $\Psi(g_{T}*g)={\cal D}(g_T)\Psi(g)$, with
${\cal D}(e_{\k},\z)=\z D(e_{\k})$, where $D(e_{\k})$ is a representation
of the group $\left\{e_{\k},\,\k\in Z\times Z\right\}\approx Z\times Z$ on
the complex numbers. For the moment, we shall use the trivial representation
$D^0(e_{\k})=1$, and later the rest of non-equivalent representations
(leading to non-equivalent quantization of \Gt) will be computed with
the help of the bad operators, as was shown in Sec. 2.1.1 for the
cylindrical H-W group.
The $T$-function condition then reads:
\be
\e^{\idhbar m\w[(1+\lambda)k_1L_1x_2 + \lambda k_2L_2x_1]}
    \Psi^{0}(\x+\L_{\k},\z)= \Psi^{0}(\x,\z) \label{H-WtConstraint}
\ee

\ni $\forall \k\in Z\times Z$. Note that the space of constrained wave
functions can be identified with the space of sections of a $U(1)$-bundle
on the torus, with Chern class $n$ and connection given by (\ref{1formT}).

Applying this constraint to the polarized wave functions (\ref{H-WtWF}) the
following restriction is obtained:
\bea
\e^{\idhbar m\w\left\{ y_2 + (1+2\lambda)n_1n_2y_1+
  [(1+\lambda)n_2^2-\lambda n_1^2](\n\cdot\J\cdot\L_{\k})\right\}
    (\n\cdot\L_{\k})}\times & & \nn \\
\e^{-\idhbar m\w\left[(1+2\lambda)n_1n_2(y_1+y_2)
    \right](\n\cdot\J\cdot\L_{\k})}
\Phi^{0}(y_2+\n\cdot\J\cdot\L_{\k}) &=& \Phi^{0}(y_2)
\eea

\ni $\forall \k\in Z\times Z$. This restriction has important consequences:
$a)$ the possible polarizations are only those given by $\n=(1,0)$ and
$\n=(0,1)$;
$b)$ the wave function is peaked at certain equally spaced values of $y_2$;
and $c)$ the parameter $\lambda$ is also quantized. From these facts it can
also
be deduced that the dimension of the representation is $n$, i.e. the
representations of the toral Heisenberg-Weyl group are finite-dimensional,
having dimension $n$, where $n$ is given by (\ref{H-WtQuant}).

Explicitly, the ``allowed" values for the coordinates are
$x_2=\frac{k}{n}L_2,\, k\in Z$ for $\n=(1,0)$
and $x_1=\frac{k}{n}L_1,\, k\in Z$ for $\n=(0,1)$.
The wave functions then turn out to be, respectively:
\bea
\Phi^{0}(x_2)=\sum_{k\in Z}a_k \delta(x_2-\frac{k}{n}L_2)
    &\hbox{for}& \n=(1,0) \label{WFt1} \\
\Phi^{0}(x_1)=\sum_{k\in Z}b_k \delta(x_1-\frac{k}{n}L_1)
    &\hbox{for}& \n=(0,1)  \label{WFt2}
\eea

\ni The coefficients $a_k$ and $b_k$ are not completely arbitrary; due to the
$T$-function
condition, which now reads $\Phi^0(x_2+k_2L_2)=\Phi^0(x_2)$ (for $\n=(1,0)$)
and $\Phi^0(x_1+k_1L_1)=\Phi^0(x_1)$ (for $\n=(0,1)$) $\forall \k\in Z\times
Z$,
they satisfy
$a_{k+n}=a_k$, and $b_{k+n}=b_k,\,\forall k\in Z$. Then, there are only
$n$ independent
coefficients, so that the dimension of the representation is $n$.
The allowed values for $\lambda$ are given by $\lambda=\frac{k}{n}\,,\, k\in
Z$,
i.e. the possible (equivalent) cocycles, or that which is the
same, the possible coboundaries are quantized. This fact can be easily
understood in terms of the generating function of the coboundary parameterized
by $\lambda$, which has
the form $\lambda x_1x_2$, or better, $\e^{\idhbar m\w\lambda x_1x_2}$. For
this function to be quasi-periodic, i.e. $\e^{\idhbar m\w\lambda
(x_1+k_1L_1)(x_2+k_2L_2)}=
\e^{i\ep\cdot\L_{\k}}\e^{\idhbar m\w\lambda x_1x_2},\forall\k\in Z\times Z$,
the quantization condition for $\lambda$ is necessary, besides the quantization
condition for $x_1$ and $x_2$.

Let us focus on the case $\n=(1,0)$ for concreteness (the case $\n=(0,1)$ is
completely analogous and in fact equivalent). Using the expression
(\ref{WFt1}) and the fact that $a_{k+n}=a_k, \forall k\in Z$, the summatory can
be regrouped, and we arrive at a rather compact form for the wave functions:
\bea
\Phi^0(x_2)&=&\sum_{k\in Z}a_k \delta(x_2-\frac{k}{n}L_2)
            =\sum_{k=0}^{n-1}\,\sum_{k_2\in Z}a_{k+nk_2}
            \delta(x_2-\frac{k+nk_2}{n}L_2) \nn \\
           &=&\sum_{k=0}^{n-1} a_k \sum_{k_2\in Z}
              \delta(x_2-\frac{k+nk_2}{n}L_2) \nn \\
           &=&\sum_{k=0}^{n-1} a_k \Lambda^0_k(x_2)
\eea

\ni where ($x_2^{(k)}\equiv x_2-\frac{k}{n}L_2$)
\be
\Lambda^0_k(x_2)\equiv\sum_{k_2\in Z}\delta(x_2^{(k)}-k_2L_2)
             = \frac{1}{L_2}\sum_{q\in Z} \e^{i2\pi q x_2^{(k)}/L_2}
\ee

\ni Therefore, the dimension of the Hilbert space ${\cal H}^0(\Gtm)$ is $n$,
since it is spanned by the functions $\Lambda^0_k(x_2),\,k=0,1,...,n-1$.

Next, we determine the subgroup $\GH$ of good
transformations (those preserving the structure of the wave
function). As in the case of the cylindrical H-W group, it is
deduced from $Ad(\Gtm)\left[e_{\kp},g\right]=
\left(0,0,\e^{\idhbar m\w(k_1L_1x_2 - k_2L_2x_1)}\right)\subset\GP\,\,
\forall \k\in Z\times Z$,
which implies $\frac{m\w}{\hbar}(k_1L_1x_2-k_2L_2x_1)=2\pi k, \,k\in Z$, and,
together with the quantization condition (\ref{H-WtQuant}) for $\w$, leads
to $\x=\frac{1}{n}\L_{\k}$. Therefore, the subgroup $\GH$ of good
transformations is the subgroup of \Gt in which the parameters $\x$ are
restricted to be $\x=\frac{1}{n}\L_{\k}$, although only a finite number of
them corresponding to
$\{\x=(\frac{k_1}{n}L_1,\frac{k_2}{n}L_2),\, k_1,k_2=0,1,...,n-1\}\,$ are
actually different, due to the $T$-function condition.
Consequently, no infinitesimal transformation (apart from that of $U(1)$)
preserves the structure of the wave function.

If we introduce the (finite) operators
$\hat{\eta}_i\equiv \e^{L_i\XR{x_i}},\, i=1,2$, in a similar way as in
Sec. 2.1.1 (although here they represent finite translations), we can write
the elements of $T$ as $e_{\k}\equiv (\hat{\eta}_1)^{k_1}(\hat{\eta}_2)^{k_2}$,
and the subgroup of good operators is:
\be
\GH =
\left\{\z(\hat{\eta}_1)^{\frac{k_1}{n}}(\hat{\eta}_2)^{\frac{k_2}{n}},\;\;
k_1,k_2\in Z,\,\,\z\in U(1) \right\} \label{IGood}
\ee

As in Sec. 2.1.1, the set of bad
operators can be interpreted as quantization-changing operators, sweeping the
space of all non-equivalent quantizations. As was proven there,
the action of a bad operator takes the wave function out of our Hilbert space
${\cal H}^0(\Gtm)$ and puts it into a different Hilbert space
${\cal H}^{\av}(\Gtm)$ corresponding to a non-equivalent
representation of $T_B$ ($=T$) parameterized by $\av$. Thus, we define the new
functions (we restrict ourselves to the $\Phi(x_2)$ part of the wave function):
\bea
\Phi^{\av}(x_2)&\equiv& \e^{\a_1\XR{x_1}+\a_2\XR{x_2}}\Phi^0(x_2)
        = \e^{i2\pi n \frac{x_2}{L_2}\frac{\a_1}{L_1}}\Phi^0(x_2+\a_2) \nn \\
       &=& \sum_{k=0}^{n-1}a_k \e^{i2\pi n \frac{x_2}{L_2}\frac{\a_1}{L_1}}
           \Lambda^0_k(x_2+\a_2)
        = \sum_{k=0}^{n-1}a_k \Lambda^{\av}_k(x_2) \label{IWF1}
\eea

\ni where
\be
\Lambda^{\av}_k(x_2)\equiv \e^{i2\pi n
\frac{x_2}{L_2}\frac{\a_1}{L_1}}\Lambda^0_k(x_2+\a_2)
     = \e^{i2\pi n \frac{x_2}{L_2}\frac{\a_1}{L_1}}\frac{1}{L_2}
       \sum_{q\in Z} \e^{i2\pi q(x_2^{(k)}+\a_2)/L_2} \label{IWF2}
\ee

\ni and the values of $\av$  are different from $\frac{1}{n}\L_{\k}$
(good transformations).

To determine the non-equivalent quantizations (i.e. the minimum range of
values of the parameters $\a_1$ and $\a_2$ that sweeps the whole set of
non-equivalent quantizations) we let the transformations of $T$ act on these
new functions and then we determine the quasi-periodicity conditions:
\bea
(\hat{\eta}_1)^{k_1} \Phi^{\av}(x_2)&=&\e^{-i2\pi n\frac{\a_2}{L_2}k_1}
                                      \Phi^{\av}(x_2) \label{IQuasiP1} \\
(\hat{\eta}_2)^{k_2} \Phi^{\av}(x_2)&=&\e^{i2\pi n\frac{\a_1}{L_1}k_2}
                                      \Phi^{\av}(x_2) \label{IQuasiP2}
\eea

\ni from which it can be deduced that $\a_1\in [0,\frac{L_1}{n})$ and
$\a_2\in [0,\frac{L_2}{n})$. This range of values is associated with
the first Brillouin zone of the reciprocal lattice, as can be checked if we
define the parameters $\epv\equiv m\w \J\cdot\av$.
It is easy to verify that the wave functions
$\{\Lambda^{\av}_k(x_2),\,k=0,1,...,n-1\}$  constitute the carrier space
(they span ${\cal H}^{\av}(\Gtm)$)
for unitary irreducible representations (parameterized by $\av$) of the
subgroup of good operators. Under these operators the wave functions transform
as:
\bea
(\hat{\eta}_1)^{k_1/n}\Lambda^{\av}_k(x_2)&=&
\e^{i2\pi(\frac{k}{n}-\frac{\a_2}{L_2})k_1}\Lambda^{\av}_k(x_2) \\
(\hat{\eta}_2)^{k_2/n}\Lambda^{\av}_k(x_2)&=&
\e^{i2\pi\frac{\a_1}{L_1}k_2}\Lambda^{\av}_{k-k_2\,{\rm mod}\,n }(x_2)
\eea

In a recent paper, \cite{GotayToro}, it is shown that, for the case $n=1$, the
symplectic manifold defined by the torus can be fully quantized, i.e. the
entire
Poisson algebra on the torus can be irreducibily represented by self-adjoint
operators acting on a Hilbert space. Here, the same result is obtained for
arbitrary integer $n$. Even more,  more operators than
those associated with classical functions (those of $T$) can be
irreducibily represented, namely
$(\hat{\eta}_i)^{\frac{k_i}{n}}, k_i\in Z, i=1,2$. To be precise, the
$n^{th}$'s roots of the classical functions can be quantized according to our
scheme. This is possible thanks to the fact that the representation defined by
the equations above is a vector representation, i.e. wave functions are really
sections of an associated vector bundle of dimension $n$ over the torus.

As in Sec. 2.1.1, we could consider the union of all the Hilbert spaces
$\cup_{\av}{\cal H}^{\av}(\Gtm)$. In this Hilbert space, the bad operators
$\XR{\x}$ are Hermitian and act irreducibily, carrying a unitary irreducible
representation of the toral H-W group, turning out to be a generalization,
for arbitrary integer $n$, of that called $kq$-representation in Solid State
Physics \cite{Zak}, where only the case $n=1$ is considered.

{\it Summarizing} the integer case, there is a continuum of non-equivalent
quantizations, corresponding
to non-equivalent representations of $T$ parameterized
by $\av$, giving rise to different quasi-periodic boundary conditions. The
value $\av=0$, corresponding to the trivial representation $D^0(e_{\k})=1$ of
$\left\{e_{\k};\,\k\in Z\times Z\right\}$, reproduces the standard periodic
boundary conditions. The wave functions are (\ref{IWF1}-\ref{IWF2}) with
quasi-periodicity conditions given by (\ref{IQuasiP1}-\ref{IQuasiP2})  and the
subgroup of good operators is (\ref{IGood}).

It should be stressed the difference between the two representations here
obtained.
On the one hand, for a fixed $\av$, the Hilbert space ${\cal H}^{\av}(\Gtm)$
carries an irreducible representation of the subgroup of good operators. In
this representation the operators $\XR{\x}$ do not preserve the Hilbert space;
they are bad operators.
On the other hand, the union of all the Hilbert spaces
$\cup_{\av}{\cal H}^{\av}(\Gtm)$ carries an irreducible representation of the
entire toral H-W group, in such a way that the operators $\XR{\x}$ are
Hermitian and the good operators act in a diagonal form.

A brief comment is now in order. Let us consider the discrete (infinite)
subgroup generated by
$\left\{(\hat{\eta}_1)^{\frac{k_1}{n}}(\hat{\eta}_2)^{\frac{k_2}{n}},\;\;
k_1,k_2\in Z \right\}$, which constitutes a principal fibre bundle with base
$Z\times Z$ and fibre $Z_n\subset U(1)$. The group algebra of this discrete
group can be proven to be (in a suitable basis) an infinite-dimensional
trigonometric algebra \cite{Fairlie}. Since $n$ is an integer, this discrete
group has a centre, which can be removed by means of the $T$-function
condition.
The quotient group is the finite group generated by
$\left\{(\hat{\eta}_1)^{\frac{k_1}{n}}(\hat{\eta}_2)^{\frac{k_2}{n}},\right.$
$\;\;$$\left. k_1,k_2=0,...,n-1 \right\}$. This finite group (which can be
seen as a finite version of the Heisenberg-Weyl group) constitutes a
principal fibre bundle with base
$Z_n \times Z_n$ and fibre $Z_n\subset U(1)$, and admits a simple matrix
representation given in Ref. \cite{Weyl} (see also \cite{Fairlie} and
\cite{Floratos}). The corresponding group algebra is the algebra of
$SU(n)\times U(1)$ for $n$ odd or $U(n/2)$ for $n$ even, in a
trigonometric basis \cite{Fairlie}. By means of this representation, the limit
$n\rightarrow \infty$ (the ``classical" limit) is particularly simple, leading
to
the algebra of infinitesimal area-preserving diffeomorphisms of a 2D-surface
(the torus, in this case). This algebra, referred to as $\w_\infty$ in the
literature, is the classical version of a variety of infinite-dimensional
algebras
called collectively $W_\infty$, of increasing interest nowadays (see
\cite{Shen}
for a review). In this sense, the subgroup of good operators $\GH$ can be seen
as the quantum version of the area-preserving diffeomorphisms of the
torus, thus constituting a realization of the $W_\infty$ algebras on the torus.

\subsubsection{Fractional quantum numbers}

We now consider the rational case, in which
$\frac{m\w L_1L_2}{2\pi\hbar}=\frac{n}{r}$. In this case
$T$ has a non-trivial characteristic subgroup, i.e. there are non-trivial
elements commuting with the whole group $T$. This is
$\GC=\{ r\L_{\k}, \k\in Z^2\}$, and  the polarization subgroup, which
must contain $\GC$, is $T_p=\GC\cup \{k\L_{\k_p}, k\in Z\}$, where
$\k_p$ is a vector the components of which are either relative prime integers,
(1,0) or (0,1). This condition is required for maximality
of the polarization subgroup, and therefore for the irreducibility of the
representation of $T$.

The $T$-function condition now reads $\Psi(g_{T_B}*g)=
{\cal D}(g_{T_B})\Psi(g)$,
where $T_B\equiv T_p\cup U(1)$ is the maximal subgroup of
compatible constraints that can be applied to the wave function, and
${\cal D}(g_{T_B})$ is a representation of $T_B$
on the complex numbers. For the moment, we shall use the
representation ${\cal D}^0(e_{T_p},\z)=\z$, which is trivial for the
elements in $T_p$. Later, the non-equivalent representations of $T_B$ will
be straightforwardly computed, as in Sec 2.2.1.
The $T_B$-function condition on the polarized
wave functions (\ref{H-WtWF}) is then:
\bea
\e^{\idhbar m\w\left\{ y_2 + (1+2\lambda)n_1n_2y_1+
    [(1+\lambda)n_2^2-\lambda n_1^2](\n\cdot\J\cdot\L_{r\k+k\k_p})
   \right\}(\n\cdot\L_{r\k+k\k_p})}\times & & \nn \\
\e^{-\idhbar m\w\left[(1+2\lambda)n_1n_2(y_1+y_2)
    \right](\n\cdot\J\cdot\L_{r\k+k\k_p})}
\Phi^0(y_2+\n\cdot\J\cdot(\L_{r\k+k\k_p})) &=& \Phi^0(y_2)
\eea

\ni $\forall k\in Z$, and $\forall \k\in Z\times Z$. As in the
integer case, the only polarization vectors $\n$ consistent with these
restrictions are $\n=(1,0)$
and $\n=(0,1)$, and the same for $\k_p$, for which the only possible values
are $\k_p=(1,0)$ and $\k_p=(0,1)$.

Let us fix the polarization to $\n=(1,0)$ for concreteness (the case $\n=(0,1)$
is completely analogous and in fact leads to an equivalent representation).
The two different choices of $\k_p$, perpendicular and parallel to $\n$,
lead to {\it non-equivalent} representations [this is a general feature
in AQG: for a given polarization in \Gt, different choices of polarization
subgroups $T_p$ in $T$ can lead to non-equivalent quantizations, even though
the polarization subgroups were equivalent from the point of view of the
subgroup $T$ itself (see \cite{Alfonso})], both with
dimension $n$ and with $\lambda$ restricted to be $\lambda=k/n,\,k\in Z$:
\begin{list}{}{ \setlength{\leftmargin}{2cm} \setlength{\rightmargin}{0cm}
                \setlength{\labelsep}{0.5cm} }
\item[\raggedright a) $\L_{\k_p}\perp \n$\hfil] i.e. $\k_p=(0,1)$, then the
wave function is
      peaked at the values $y_2=x_2=\frac{k}{n}L_2,\, k\in Z$, satisfies
      $\Phi^0_\perp(x_2+k_2L_2)=\Phi^0_\perp(x_2)$,  and has the form
\be
    \Phi^0_\perp(x_2)=\sum_{k=0}^{n-1}a_k\Lambda_k^0(x_2)
\ee
     where $\Lambda_k^0(x_2)$ is defined as in Sec 2.2.1, and the subgroup of
     good transformations is $\GH=\{\frac{r}{n}\L_{\k},\,\k\in Z\times Z\}
     \cup\{\frac{k}{n}L_2,\,k\in Z\}$, although only a finite subgroup of them
     are distinct:
\be
 \GH^\perp =\left\{(\hat{\eta}_1)^{r\frac{k_1}{n}},\,
(\hat{\eta}_2)^{\frac{k_2}{n}},\;\;
k_1,k_2=0,...,n-1 \right\} \label{FGoodperp}
\ee
\item[\raggedright b) $\L_{\k_p}\parallel \n$\hfil] i.e. $\k_p=(1,0)$, then the
wave function
       is peaked at the values
       $y_2=x_2=k\frac{r}{n}L_2,\, k=0,1,...n-1$, satisfies
       $\Phi^0_\parallel(x_2+rk_2L_2)=\Phi^0_\parallel(x_2)$, and has the form
\be
  \Phi^0_\parallel(x_2)=\sum_{k=0}^{n-1}a_k\Lambda^{r,0}_k(x_2)
\ee
     where $\Lambda^{r,0}_k(x_2)\equiv
      \frac{1}{rL_2}\sum_{q\in Z} \e^{i2\pi q x_2^{r,(k)}/(rL_2)}$, with
     $x_2^{r,(k)}\equiv x_2-\frac{k}{n}rL_2$, and the subgroup
     of good transformations is $\GH=\{\frac{r}{n}\L_{\k},\,
     \k\in Z\times Z\}\cup\{\frac{k}{n}L_1,\,k\in Z\}$. Again, only a finite
     subgroup of them are distinct:
\be
 \GH^\parallel =\left\{(\hat{\eta}_1)^{\frac{k_1}{n}},\,
(\hat{\eta}_2)^{r\frac{k_2}{n}},\;\;
k_1,k_2=0,...,n-1 \right\} \label{FGoodparallel}
\ee
\end{list}

As in Sec. 2.1.1, we can compute the non-equivalent representations by
applying the whole set of bad operators to the wave functions. We proceed as
in the integer case (Sec 2.2.1) and obtain:
\begin{list}{}{ \setlength{\leftmargin}{2cm} \setlength{\rightmargin}{0cm}
                \setlength{\labelsep}{0.5cm} }
\item[\raggedright a) $\L_{\k_p}\perp \n$\hfil] the wave functions have the
form
\be
    \Phi^{\av_p}_\perp(x_2)=\sum_{k=0}^{n-1}a_k\Lambda_k^{\av_p}(x_2)
           \label{FWFperp}
\ee
      with $\Lambda_k^{\av_p}(x_2)\equiv \e^{i2\pi\frac{n}{r}
            \frac{\a_{p1}}{L_1}\frac{x_2}{L_2}}\Lambda_k^{0}(x_2)$. They
     satisfy
\bea
(\hat{\eta}_1)^{rk_1}\Phi^{\av_p}_\perp(x_2)&=&
             \e^{-i2\pi n\frac{\a_{p2}}{L_2}k_1}\Phi^{\av_p}_\perp(x_2)
             \label{FQuasiPperp1}\\
(\hat{\eta}_2)^{k_2}\Phi^{\av_p}_\perp(x_2)&=&
             \e^{i2\pi n\frac{\a_{p1}}{rL_1}k_2}\Phi^{\av_p}_\perp(x_2)
             \label{FQuasiPperp2}
\eea
       with $\a_{p1}\in [0,r\frac{L_1}{n}),\,\a_{p2}\in [0,\frac{L_2}{n})$.
\item[\raggedright b) $\L_{\k_p}\parallel \n$\hfil] the wave functions have the
form
\be
  \Phi^{\bv_p}_\parallel(x_2)=\sum_{k=0}^{n-1}a_k\Lambda^{r,\bv_p}_k(x_2)
           \label{FWFparallel}
\ee
     with $\Lambda_k^{r,\bv_p}(x_2)\equiv \e^{i2\pi\frac{n}{r}
            \frac{\b_{p1}}{L_1}\frac{x_2}{L_2}}\Lambda_k^{r,0}(x_2)$. They
     satisfy
\bea
(\hat{\eta}_1)^{k_1}\Phi^{\bv_p}_\parallel(x_2)&=&
             \e^{-i2\pi n\frac{\b_{p2}}{rL_2}k_1}\Phi^{\av_p}_\parallel(x_2)
             \label{FQuasiPparallel1} \\
(\hat{\eta}_2)^{rk_2}\Phi^{\bv_p}_\parallel(x_2)&=&
             \e^{i2\pi n\frac{\b_{p1}}{L_1}k_2}\Phi^{\bv_p}_\parallel(x_2)
             \label{FQuasiPparallel2}
\eea
       with $\b_{p1}\in [0,\frac{L_1}{n}),\,\b_{p2}\in [0,r\frac{L_2}{n})$.
\end{list}

\ni It is easy to verify that the wave functions
$\{\Lambda^{\av_p}_k(x_2),\,k=0,1,...,n-1\}$ and
$\{\Lambda^{r,\bv_p}_k(x_2),\,k=0,1,...,n-1\}$ constitute the carrier spaces
for unitary irreducible representations (parameterized by $\av$ and $\bv$,
respectively) of the
subgroup of good operators. Under these operators they transform as:
\begin{list}{}{ \setlength{\leftmargin}{2cm} \setlength{\rightmargin}{0cm}
                \setlength{\labelsep}{0.5cm} }
\item[\raggedright a) $\L_{\k_p}\perp \n$\hfil]
     \bea
     (\hat{\eta}_1)^{rk_1/n}\Lambda^{\av_p}_k(x_2)&=&
     \e^{i2\pi(\frac{k}{n}-\frac{\a_{p2}}{L_2})}\Lambda^{\av_p}_k(x_2) \\
     (\hat{\eta}_2)^{k_2/n}\Lambda^{\av_p}_k(x_2)&=&
     \e^{i2\pi\frac{\a_1}{rL_1}k_2}\Lambda^{\av_p}_{k-k_2\,{\rm mod}\,n }(x_2)
     \eea
\item[\raggedright b) $\L_{\k_p}\parallel \n$\hfil]
     \bea
     (\hat{\eta}_1)^{k_1/n}\Lambda^{r,\bv_p}_k(x_2)&=&
     \e^{i2\pi(\frac{k}{n}-\frac{\b_{p2}}{rL_2})}\Lambda^{r,\bv_p}_k(x_2) \\
     (\hat{\eta}_2)^{rk_2/n}\Lambda^{r,\bv_p}_k(x_2)&=&
     \e^{i2\pi\frac{\b_{p1}}{L_1}k_2}\Lambda^{r,\bv_p}_{k-k_2\,{\rm mod}\,n
}(x_2)
     \eea
\end{list}

It should be noted that although $\frac{m\w L_1L_2}{2\pi\hbar}=\frac{n}{r}$,
the dimension
of the representations is $n$, and $\lambda=k/n,\,k\in Z$, as in the integer
case (even more, in the case $\L_{\k_p}\perp \n$ the wave functions coincide);
the difference is found in the subgroups of good operators, which, although
isomorphic, differ in the specific values of the transformations. This
representation can be reinterpreted as mimicking a torus
$r$ times greater in one direction
(determined by the orthogonal vector to $\k_p$), i.e., the
area of the effective torus is $rL_1L_2$, and therefore
$\frac{m\w (r L_1L_2)}{2\pi\hbar}=n$. Thus, the same results as in the integer
case now apply, although changing $L_2$ by $rL_2$ if $\k_p=(1,0)$ or
$L_1$ by $rL_1$ if $\k_p=(0,1)$.

{\it Summarizing} the fractional case, there are two continua of
non-equivalent quantizations, according to the choices $\L_{\k_p}\perp \n$ and
$\L_{\k_p}\parallel \n$, parameterized by $\av_p$ and $\bv_p$, respectively.
The wave functions are given by (\ref{FWFperp}) and (\ref{FWFparallel}),
satisfying quasi-periodicity conditions given by
(\ref{FQuasiPperp1}-\ref{FQuasiPperp2}) and
(\ref{FQuasiPparallel1}-\ref{FQuasiPparallel2}), respectively. The subgroups
of good operators are given by (\ref{FGoodperp}) and (\ref{FGoodparallel}),
respectively.

\ni {\it Associated $r$-vector bundle:}
If we act on the wave functions with the bad operators of $T$ (i.e. those
operators of $T$ which are not in $T_B$) the resulting
wave functions lie in a different Hilbert space belonging to a different
quantization. However, as these operators are finite and their $r^{th}$ power
are good operators, these new wave functions transform among each other under
the action of the subgroup $T_{\rm bad}$, defined as the set of bad operators
of $T$
and the identity. Therefore, constructing the vector space spanned by these $r$
functions ($T_{\rm bad}$ has $r$ elements), we obtain an $r$-dimensional,
unitary irreducible representation of the group $T$ as a whole, including the
bad operators.
Explicitly:
\begin{list}{}{ \setlength{\leftmargin}{2cm} \setlength{\rightmargin}{0cm}
                \setlength{\labelsep}{0.5cm} }
\item[\raggedright a) $\L_{\k_p}\perp \n$\hfil] we define
     \be
     \Lambda^{\av_p}_{k,j}(x_2)\equiv(\hat{\eta}_1)^{j}\Lambda^{\av_p}_k(x_2)=
     \e^{i2\pi\frac{n}{r}\frac{x_2}{L_2}j}\Lambda^{\av_p}_k(x_2)
     \ee
  for $j=0,1,...,r-1$, where they satisfy:
    \be
     (\hat{\eta}_1)^{j'}\Lambda^{\av_p}_{k,j}(x_2)=
        \e^{-i2\pi n\frac{\a_{p2}}{L_2}(j+j'\,{\rm div}\, r)}
        \Lambda^{\av_p}_{k,j+j'\,{\rm mod}\, r}(x_2)
    \ee
    for $j,j'=0,1,...,r-1$.
\item[\raggedright b) $\L_{\k_p}\parallel \n$\hfil] we define
     \be
\Lambda^{r,\av_p}_{k,j}(x_2)\equiv(\hat{\eta}_2)^{j}\Lambda^{r,\av_p}_k(x_2)=
     \Lambda^{r,\av_p}_k(x_2+jL_2) \\
     \ee
     for $j=0,1,...,r-1$, satisfying:
    \be
     (\hat{\eta}_2 )^{j'}\Lambda^{r,\av_p}_{k,j}(x_2)=
        \e^{i2\pi n\frac{\a_{p1}}{L_1}(j+j'\, {\rm div}\, r)}
        \Lambda^{r,\av_p}_{k,j+j'\,{\rm mod}\, r}(x_2)
    \ee
  for $j,j'=0,1,...,r-1$.
\end{list}

This construction can be viewed as the $r$-dimensional vector bundle associated
with the principal bundle \Gt, which has structure group $T$. The $r$-component
wave functions are sections of this associated vector bundle.

As stated before, AQG generalizes Geometric Quantization in some respects,
in particular in that which concerns (topologic) quantum numbers. The
fractional value $\frac{m\w L_1L_2}{2\pi\hbar}=\frac{n}{r}$ generalizes the
integer
class of the standard symplectic form (the Chern class of the line bundle).
The geometric quantization of a symplectic manifold with ``fractional class"
$\frac{n}{r}$ would have led to r-valued wave functions (as opposed to
single-valued). Eventually, this
trouble could have been circumvented by replacing the usual line bundle by a
complex
vector bundle $E$ of rank $r$ and Chern class $n$, as constructed before.

The comments at the end of Sec. 2.2.2 concernig the generalized
$kq$-representation can be translated to the $r$-bundle structure associated
with the fractional case.

\subsubsection{Irrational case}

Finally, and for the sake of thoroughness, let us briefly comment on the case
in which
$\rho\equiv\frac{m\w L_1L_2}{2\pi\hbar}$ is an irrational number. In this case
the characteristic group is trivial, and $T_B=T_p\cup U(1)$, with
$T_p=\{k\L_{\k_p}, k\in Z\}$ only. As before, it can be proven that the
only possible polarization vectors are $\n=(1,0)$ and $\n=(0,1)$.
Moreover, the only consistent choice of polarizations $T_p$ in $T$ are also
$\k_p=(1,0)$ and $\k_p=(0,1)$. No restriction for $\lambda$ appears in this
case, and the structure of $T_B$-function condition closely
resembles that of the case of the cylindrical H-W group: the wave
functions are either peaked at an infinite series of equally spaced values of
$y_2$ if
$\k_p\parallel \n$ (as in the momentum space representation in the cylindrical
H-W group),
or quasi-periodic if $\k_p\perp \n$ (as in the
configuration space representation in the cylindrical H-W group). In both cases
the
non-equivalent representations are labelled by
$\ep\in[0,\frac{2\pi\hbar}{|\L_{\k_p}|})$.
The representations are therefore infinite dimensional, and the subgroup of
good operators is given by $\GH=\{\frac{1}{\rho}\L_{\k}, \k\in Z\times Z\}\cup
\{\alpha\L_{\k_p}, \alpha\in R\}$. Consequently, besides the discrete
transformations in
$\x=\frac{1}{\rho}\L_{\k}$, the infinitesimal operator
$\L_{\k_p}\cdot\XR{\x}$ is also a
good  operator, that is, arbitrary translations in the direction of $\L_{\k_p}$
are good transformations. Note that, $\rho$ being an irrational number,
$\frac{1}{\rho}\L_{\k}$ never reaches a point of the lattice defined by
$\L_{\k}$, although it
fills the corresponding torus densely when varying $\k\in Z\times Z$.

Therefore, in this case, the subgroups  $x_1=0$ and $x_2=0$ (the classical
circumferences),  are represented faithfully, as in the case of the cylindrical
H-W group, but the rest of the
group is not faithfully represented, nor are even the points of the lattice
(the group $T$). In
particular, for the infinite-order operators $\hat{\eta}_1,\hat{\eta}_2$,
defined as in Sec 2.1.1 for the directions $x_1,x_2$, only one is a
good operator (the one in the direction of $\L_{\k_p}$), the other being a
bad operator. Consequently, we cannot represent the toral H-W
group faithfully for irrational values of $\rho$.

\section{Free Galilean particle on the circumference}

Let us apply the results obtained in the last section to the simple
example of the free particle moving on the circumference.

We can study this problem easily by simply  adding the temporal evolution to
the results obtained in Sec. 2.1 (for the group law, vector fields,
polarizations,
Schr\"odinger equation, etc., see \cite{Position} and references therein),
without affecting the main conclusions of that section. The main new features
are the introduction of a new operator $\E$ associated with the temporal
evolution
and the fact that, by using the Schr\"odinger equation, this operator can be
written in
terms of the momentum operator as $\frac{1}{2m}\P^2$. Since $\P$ is a good
operator, $\E$ proves
also to be a good operator. A common set of eigenfunctions is given by
\bea
\phi^\ep_n(x,t)&=&\e^{-\idhbar\frac{1}{2m}(\ep+\frac{2\pi\hbar}{L}n)^2 t}
     \e^{\idhbar(\ep+\frac{2\pi\hbar}{L}n)x} \nn \\
\E\phi^\ep_n&=& \frac{1}{2m}(\ep+\frac{2\pi\hbar}{L}n)^2\phi^\ep_n \\
\P\phi^\ep_n&=& (\ep+\frac{2\pi\hbar}{L}n)\phi^\ep_n \nn
\eea

\ni where $n\in Z$. Note that for $\ep=0$ the states $n$ and $-n$ have
the same energy, which means that all the energy eigenstates except for the
vacuum are degenerate. For any other value of $\ep$, the
states $n$ and $-(n+2\ep\frac{L}{2\pi\hbar})$ have the same energy,
but $-(n+2\ep\frac{L}{2\pi\hbar})$ is
an eigenstate only if $2\ep\frac{L}{2\pi\hbar}\in Z$, i.e.
$\ep\frac{L}{2\pi\hbar}$ is integer or half-integer. This means that,
in general, there is no degeneracy for any value of $\ep$ except for the
integer values, in which case all the eigenstates are doubly degenerate
except for the
vacuum, and half-integers, for which all the eigenstates, including the
vacuum, are doubly degenerate. The phenomenon of degenerate ground state in
this
simple model  parallels $\theta$-vacuum phenomenon in Yang-Mills field
theories \cite{Asorey}.

The feature of non-equivalent quantizations can be reproduced (in an
equivalent  way, indeed) by the introduction of an extra coboundary in
(\ref{H-WcGLaw}) (more precisely, in its counterpart when the temporal
evolution is added; see \cite{Position}) generated by the function
$\ep x$, i.e. a multiplicative factor of the form
$\e^{\idhbar \ep \frac{p'}{m}t}$ in the $\z\in U(1)$ composition
law [We remind that $x''=x'+x+\frac{p'}{m}t$ is the composition law
for $x$ when the temporal evolution is added]. In the case of the free Galilean
particle on the real line, the only consequence of this term is the appearance
of a total derivative in the quantization 1-form $\Theta$
(or, what is the same, in the Lagrangian), leading thus to equivalent
(classical
and quantum) theories, as expected from the fact that $\ep \frac{p'}{m}t$ is
a coboundary. The situation is quite different when the particle is on the
circumference: the generating function $\ep x$, or better $\e^{\idhbar\ep x}$,
is not single-valuate on the circumference unless
$\ep=\frac{2\pi\hbar}{L}k,\,k\in Z$. As a consequence, two cocycles differing
in a coboundary generated by $\ep x$ (and therefore leading to equivalent
theories on the real line) lead to non-equivalent theories on the
circumference if $\ep\neq\frac{2\pi\hbar}{L}k,\,k\in Z$. This process of
creation of non-trivial cohomology  closely resembles the appearance of
cohomology under the process of group contraction, as in the case of the
Poincar\'e group, in which a certain class of coboundaries (generated by a
linear function in time) become true cocycles in the
$c\rightarrow\infty$ limit since their generating function goes to infinity in
this limit.

Another interesting way of interpreting the feature of non-equivalent
quantizations parameterized by $\ep$, at least in the case of charged
particles,
is as an Aharonov-Bohm-like effect. The
different quantizations can be carried out physically by producing (externally,
with the help of a solenoid) a magnetic
flux $\Phi$ through the circumference, in a way that the particle does not feel
the
magnetic field, but rather the vector potential only. Under these
circumstances, the
effect of the vector potential is the same as that of a boost, leading to
non-equivalent quantizations depending on the flux through the
circumference, in such a way that $\ep=e\Phi/c$. An interesting physical
application is that of a superconducting ring
threaded by a magnetic flux, where by Meissner effect the magnetic flux is
pulled out of the interior region of the superconducting ring, and therefore
the
magnetic field is effectively zero and only the vector potential is relevant
(Aharonov-Bohm effect). If the flux is [in this case the effective
electric charge is $e^*=2e$ because electrons form Cooper pairs]
$ k\Phi_0,\,k\in Z$, where $\Phi_0\equiv hc/e^*$ is the quantum unit of flux,
there is no net current in the superconducting
ring, but for any other value of the flux there is a net current which has
the form given in Figure 1.

\begin{figure}[p]
\epsfxsize=9.0cm
\centerline{\epsfbox{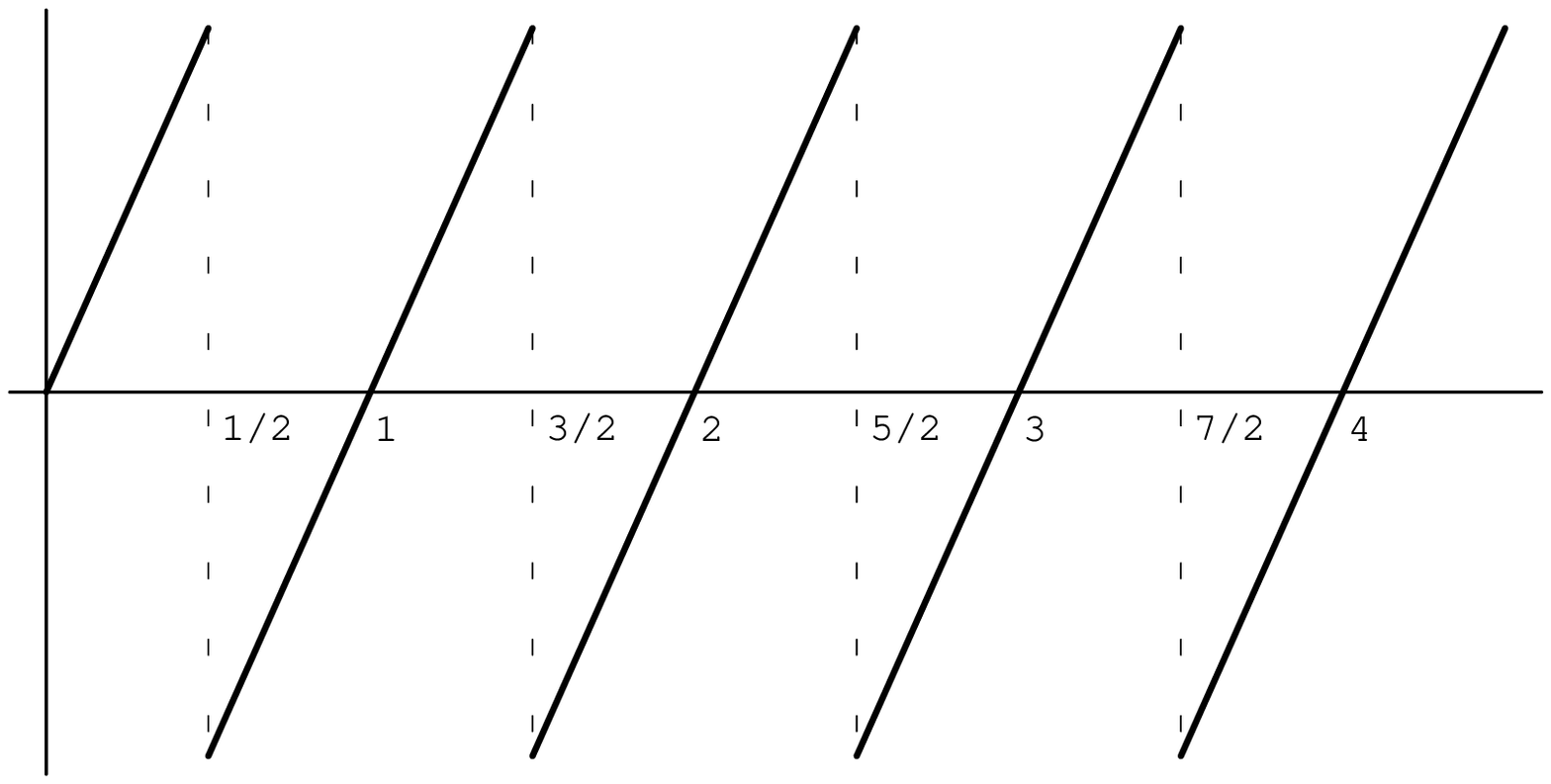}}
\caption{Net current in the superconducting ring against $\Phi/\Phi_0$.}
\end{figure}

Note that for half-integer values of $k$, the net current has no definite sign,
as a consequence, precisely, of the double degeneracy of states, in such a way
that states with opposite signs of velocity have the same energy and therefore
there is no energy cost to pass from one to the other.

\subsection{Failure of the Ehrenfest theorem}

As mentioned in the introduction, the most common problem appearing in
systems with topologically non-trivial configuration space is the failure
of Ehrenfest theorem for certain operators (``anomalous" operators)
\cite{Esteve}. Ehrenfest theorem asserts that the expectation values of quantum
operators follow classical equations of motion:
\be
\frac{d}{dt}<\hat{A}> =\idhbar <\left[\hat{H},\hat{A}\right]>
\label{Ehrenfest}
\ee

\ni In the framework of AQG, this is a natural consequence of the
appearance of bad operators, those which do not preserve the Hilbert space
of wave functions that verify the $T$-function condition.

 In Ref. \cite{Esteve} it is claimed that when the operator $\hat{A}$ does
not keep invariant the domain of $\hat{H}$, then an extra term appears in
the r.h.s. of (\ref{Ehrenfest}), which is interpreted as an anomaly.
In the language of AQG, we would say that $\hat{A}$ is a bad operator, so that
neither the left-hand side nor the right-hand side of (\ref{Ehrenfest}) would
make
sense, since the operator $\hat{A}$ takes the wave function off the Hilbert
space where the Hamiltonian $\hat{H}$ is self-adjoint (of course $\hat{H}$ is
a good operator; otherwise the temporal evolution would take the physical
states
off the Hilbert space, and the system would have no physical meaning). The
appearance of the ``anomalous" term violating the Ehrenfest theorem is a
consequence of this fact.

Returning to the free Galilean particle on the circumference, the Ehrenfest
theorem will fail for the position operator, which is a bad operator and
therefore Eq. (\ref{Ehrenfest}) makes no sense in this regard.

In conclusion, whenever there are bad operators in the theory, the Ehrenfest
theorem will fail for each of these operators and, in general, any
expectation value involving these operators will be ill-defined, giving
extra terms that can eventually be interpreted as topologic anomalies.

\section{Charged particle in a homogeneous magnetic field on the torus}

Now, we shall consider the most interesting problem of a charged particle
moving on a torus in the presence of a homogeneous magnetic field. This
problem is related to the Schwinger model \cite{Manton}, and has important
applications in the Quantum Hall effect \cite{Hall1,Hall2,Thouless}. The
magnetic field is
perpendicular to the torus surface, and the total flux is quantized (as we
shall see),
much in the same manner the Dirac monopole charge is
quantized \cite{Wu-Yang}. The actual connection of this system with
Quantum Hall Effect is based on the fact that the wave function of the complete
system factorizes in a relative-coordinate dependent term (which includes
interactions) and a centre of mass dependent term, which behaves essentially
as a particle in a transverse homogeneous magnetic field, and on the
effective
topology of the experimental device in the latter system; the topology of the
(semiconductor) strip along with the current and voltage leads is that of a
punctured torus \cite{Thouless}.

Firstly, we shall study the planar case, i.e. the charged particle on the
plane, to clarify the meaning of the different magnitudes appearing in the
problem, and to obtain a proper parameterization of the system.

\subsection{Charged particle in a homogeneous magnetic field}

The movement of a charged particle in a homogeneous magnetic field
can be factorized into a 2-dimensional problem (on the plane normal
to the magnetic field) times a free movement in the direction
of the magnetic field. Thus, we restrict ourselves
to a 2-dimensional system characterized by a non-zero commutator
between the translation generators, $[\XL{x^1},\XL{x^2}]=im\wc/\hbar$, where
$\wc$ is
the cyclotron frequency, $\omega_c=\frac{q{\cal H}}{mc}$,
${\cal H}$ the magnetic field strength and $q$ the particle electric charge
\cite{Landau,Cohen-Tannoudji}.

We have to build up a group law for this system, which must be a deformation
of the Galilean group law (in two dimensions) due to the non-zero commutator
between the translation generators. In fact, the Galilean group does not admit
any central extension giving rise to $[\XL{x^1},\XL{x^2}]=im\wc/\hbar$, and a
deformation of the non-extended algebra is required:
$\left[X^L_{t},X^L_{\x}\right] =  \Bm \J\cdot X^L_{\x}$. We then arrive at the
following Lie algebra as the quantum symmetry for our system:
\bea
\left[\XL{t},\XL{\x}\right] &=& \Bm \J\cdot\XL{\x} \nn \\
\left[\XL{t},\XL{\p}\right] &=&-\frac{1}{m} \XL{\x}  \\
\left[\XL{x^i},\XL{p^j}\right] &=& \frac{\delta_{ij}}{\hbar}\Xi \nn \\
\left[\XL{x^i},\XL{x^j}\right] &=& \frac{m\wc}{\hbar}\epsilon_{ij}\Xi \nn
\eea

A group law for this centrally extended Lie algebra becomes:
\bea
t'' &=& t' + t \nn \\
\x\,''&=& \x + \Mi\cdot\x\,' + \frac{1}{m\wc}\left(\Ni + \J\right)\cdot\p\,'
\nn \\
\p\,''&=& \p\,'+ \p \label{BGLaw} \\
\z''&=& \z'\z \e^{\idhbar\xi(g',g)} \nn
\eea

\ni where the cocycle is given by:
\be
\xi(g',g) = \medio\left\{m\wc\x\,'\cdot\N\cdot\x - \p\,'\cdot\M\cdot\x +
\x\,'\cdot\M\cdot\p +
            \frac{1}{m\wc}\p\,'\cdot\left(\N - \J \right)\cdot\p \right\}
\label{Bcocycle}
\ee

\ni The $2\times 2$ matrices are given by $\M \equiv \cos\Bmt\, \I -
\sin\Bmt\, \J\,$,
$\N\equiv  \sin\Bmt\, \I + \cos\Bmt\, \J$, $\M$ and $\N$ being orthogonal, and
$\J_{ij}\equiv\epsilon_{ij}$, $\epsilon_{12}=1$. We have not taken into
account the rotations, since they do not play any dynamical r\^ole, although
they are of interest in that, when considered on the torus, they
represent a very simple example of a local (in the strict
mathematical sense) symmetry of the equation of motion which cannot
be realized globally.

The left and right invariants vector fields are easily deduced from the group
law:
\bea
\XL{t}  & = & \parcial{t} + \frac{\p}{m}\cdot \parcial{\x}
                                  -\Bm\x\cdot\J\cdot\parcial{\x} \nn \\
\XL{\x} & = & \parcial{\x} -\frac{1}{2\hbar}\left[\p+m\wc\J\cdot\x\right]\,\Xi
               \label{XLB} \\
\XL{\p} & = & \parcial{\p} + \frac{\x}{2\hbar}\,\Xi\nn   \\
\XL{\z} & = & i\z\parcial{\z} \equiv \Xi \nn \\
              &   &  \nn \\
              &   &  \nn \\
\XR{t}  & = & \parcial{t}  \nn \\
\XR{\x} & = & \M\cdot\parcial{\x} +
             \frac{1}{2\hbar}\left[\M\cdot\p+m\wc\N\cdot\x\right]\,\Xi
\label{XRB} \\
\XR{\p} & = & \parcial{\p} + \frac{1}{m\wc}\left(\N-\J\right)\cdot\parcial{\x}
      - \frac{1}{2\hbar}\left[\M\cdot\x -
           \frac{1}{m\wc}\left(\N-\J\right)\cdot\p\right]\,\Xi \nn \\
\XR{\z} & = & i\z\parcial{\z} \equiv \Xi \nn
\eea

\ni and from (\ref{XLB}) the quantization 1-form is computed:
\be
\Theta = \medio\left[\p\cdot d\x-\x\cdot d\p - m\wc\x\cdot\J\cdot d\x\right] -
         \left[\frac{p^2}{2m} + \Bm\p\cdot\J\cdot\x +
\frac{m\wc^2}{2}\x^2\right]dt +
         \hbar\frac{d\z}{i\z}
\ee

\ni  the characteristic module of which is $\Gthetam=<\XL{t}>$. From this, the
classical equations of motion are written:
\bea
\p &=&\vec{P} \nn \\
\x &=&\Mi\cdot\ro + \frac{1}{m\wc}\J\cdot\vec{P}
\eea

\ni where $\vec{P}$ and $\ro$ are arbitrary constant vectors, parameterizing
the (classical) solution manifold. With the aid of the constant $\wc$, we may
introduce $\R\equiv\frac{1}{m\wc}\J\cdot\vec{P}$, so that the second line of
the
equation above reads $\x=\Mi\cdot\ro + \R$, i.e. the classical trajectories are
circumferences centred at $\R$, with radius $|\ro|$.

The Noether invariants, in terms of the constants $\ro$ and $\R$, are:
\bea
\iXR{t} &=& \frac{m\wc^2}{2}\ro{}^2\equiv H \nn \\
\iXR{\x} &=& m\wc\J\cdot\ro \\
\iXR{\p} &=& -(\ro + \R) \equiv \xo  \nn
\eea

\ni where $H$ is the classical energy of the system. It should be noted that
the energy
depends only on the radius $|\ro|$ of the circumference, and not on the
position
$\R$ of its centre, as corresponds to a system with translational
invariance [The system possesses translational invariance in the more
conventional sense (the magnetic field is homogeneous) although the
translation generator $\XR{\x}$ does not commute with the Hamiltonian $\XR{t}$.
In fact, as we shall see later, there exists a
translation generator in the Lie algebra (the magnetic translations) which
commutes with the Hamiltonian generator].

To obtain the representation in configuration space, we need to impose
polarization conditions similar to those of the Galilean case \cite{Position}:

\be
{\cal P}^{HO}=<\XL{\p},\XL{t}-\frac{i\hbar}{2m}\left(\XL{\x}\right)^2 >
\ee

\ni Solving the polarization equations we obtain for the wave functions
the general form:
\be
\Psi = \z \e^{-\frac{i}{2\hbar}\x\cdot\p}\Phi(\x,t) \label{BWF}
\ee

\ni where $\Phi(\x,t)$ satisfies the Schr\"odinger equation
\be
i\hbar \parcial{t}\Phi = \left\{- \frac{\hbar^2}{2m} \vec{\nabla}^2
+ i\hbar\frac{\wc}{2}\x\cdot\J\cdot\vec{\nabla} + \frac{m\wc^2}{8}\x^2\right\}
\Phi
\label{BSEQN}
\ee

The quantum operators are:
\bea
\E \Psi & = & i\hbar\parcial{t}\Psi =
       \z \e^{-\frac{i}{2\hbar}
\p\cdot\x}\left[-\frac{\hbar^2}{2m}\vec{\nabla}^2+
   i\hbar\frac{\wc}{2}\x\cdot\J\cdot\vec{\nabla} + \frac{m\wc^2}{8}\x^2
\right]\Phi(\x,t) \nn \\
\Pv \Psi & = & \z \e^{-\frac{i}{2\hbar} \p\cdot\x}
\left[-i\hbar\M\cdot\vec{\nabla}
       + \frac{m\wc}{2}\N\cdot\x \right]\Phi(\x,t) \label{BOConf}   \\
\Xv \Psi & = & \z \e^{-\frac{i}{2\hbar} \p\cdot\x}
           \left[\medio\left(\M + \I\right)\cdot\x +
  \frac{i\hbar}{m\wc}\left(\N-\J\right)\cdot\vec{\nabla}\right]\Phi(\x,t)  \nn
\eea

Instead of proceeding further and solving the Schr\"odinger equation
explicitly,
we shall perform a change of variables which
will clarify the meaning of the different magnitudes entering the theory and
which will make facilitate the accomplishment of AQG in the next subsection.
If we define $\r\equiv\Mi\cdot\ro$, we can easily rewrite the group law
(\ref{BGLaw}) and (\ref{Bcocycle}) in terms of $\r$ and $\R$:
\bea
t'' &=& t' + t \nn \\
\r{\:}'' &=& \r + \Mi\cdot\r{\:}' \nn  \\
\R{}'' &=& \R{}' + \R    \\
\z'' &=& \z'\z \e^{\idhbar m\wc\left[\medio\r{\:}'\cdot\N\cdot\r
             -\left((1+\lambda)R_1'R_2-\lambda R_2' R_1\right)\right]} \nn
\eea

\ni where we have added the coboundary generated by
$-m\wc(\medio+\lambda)R_1R_2$ to accommodate the cocycle, in its
$\R$-dependent term, to the expression of Sec. 2.2 (except for a global minus
sign).

\ni From this group law we can compute again the left- and right-invariant
vector fields:
\be
\ba{lcl}
\XL{t}  & = & \parcial{t}  -\Bm\r\cdot\J\cdot\parcial{\r}  \\
\XL{\r} & = & \parcial{\r} -\frac{m\wc}{2\hbar}\J\cdot\r\,\Xi
                \\
\XL{R_1} & = & \parcial{R_1} - \frac{\lambda}{\hbar}m\wc R_2\,\Xi   \\
\XL{R_2} & = & \parcial{R_2} - \frac{1+\lambda}{\hbar}m\wc R_1\,\Xi   \\
\XL{\z} & = & i\z\parcial{\z} \equiv \Xi
\ea \,\,\,\,\,\,\,
\ba{lcl}
\XR{t}  & = & \parcial{t}   \\
\XR{\r} & = & \M\cdot\parcial{\r} +
             \frac{m\wc}{2\hbar}\N\cdot\r\,\Xi \label{XRBn} \\
\XR{R_1} & = & \parcial{R_1} - \frac{1+\lambda}{\hbar}m\wc R_2\,\Xi  \\
\XR{R_2} & = & \parcial{R_2} - \frac{\lambda}{\hbar}m\wc R_1\,\Xi  \\
\XR{\z} & = & i\z\parcial{\z} \equiv \Xi
\ea \label{XLRBn}
\ee

\ni and the commutation relations are now:
\bea
\left[\XL{t},\XL{\r}\right] &=& \Bm \J\cdot\XL{\r} \nn \\
\left[\XL{t},\XL{\R}\right] &=& 0 \nn  \\
\left[\XL{r^i},\XL{r^j}\right] &=& \frac{m\wc}{\hbar}\epsilon_{ij}\Xi
\label{CommB}\\
\left[\XL{R^i},\XL{R^j}\right] &=& - \frac{m\wc }{\hbar}\epsilon_{ij}\Xi\nn\\
\left[\XL{\r},\XL{\R}\right] &=& 0 \nn
\eea


A glance at the algebra (\ref{CommB}) reveals that it
is the central extension of the direct
sum of the harmonic
oscillator algebra and the Heisenberg algebra. Consequently, the wave function
factorizes into a harmonic oscillator
wave function (depending on $t$ and $\r$) times a function of
$\R$, and the energy spectrum coincides with that of the harmonic
oscillator, the degeneracy being infinite due to the Heisenberg-Weyl symmetry,
which in the
plane has only infinite-dimensional unitary irreducible representations.

We are interested in a configuration-space representation, so that
a second-order polarization is needed. This is found to be:
\be
{\cal P}^{HO}=<\XL{p},\XL{t}-\frac{i\hbar}{2m}\left(\XL{\r}\right)^2,
\n\cdot\XL{\R},\np\cdot\XL{\r} >
\ee

\ni where $\n$ and $\np$ are arbitrary unit vectors. They can be chosen
to be $(1,0)$ or $(0,1)$, for instance.

Imposing these polarization conditions to the wave functions, we obtain
the general form:
\be
\Psi= \z \e^{-\idhbar m\w[ (\lambda n_1^2-(1+\lambda)n_2^2)y_1y_2 +
     (\lambda + \medio)n_1n_2 y_1^2]} \Phi(y_2)
    \e^{\frac{im\wc}{2\hbar}\kappa_2\kappa_1}
    \Omega(\kappa_2,t)\label{BWFr-R}
\ee

\ni where $y_1\equiv \n\cdot\R,\, y_2\equiv \n\cdot\J\cdot\R,\,
\kappa_1\equiv\np\cdot\r,\,\kappa_2\equiv\np\cdot\J\cdot\r,\,\,\Phi(y_2)$ is
an arbitrary function and $\Theta(\kappa_2,t)$ satisfies the Schr\"odinger
equation:
\be
i\hbar\parcial{t}\Omega(\kappa_2,t)=\left[-\frac{\hbar^2}{2m}
      \vec{\nabla}_{\kappa_2}^2
  + \frac{m\wc^2}{2}\kappa_2^2\right]\Omega(\kappa_2,t)
\ee

\ni This is nothing other than the Schr\"odinger equation for the
harmonic oscillator, so that the  solutions are
\be
\Omega(\kappa_2,t)=\sum_n A_n \e^{-i(n+\medio)\wc t}
   \e^{-\frac{m\wc}{2\hbar}\kappa_2^2}H_n(\sqrt{\frac{m\wc}{\hbar}}\kappa_2)
\ee

\ni where $H_n$ are the Hermite polynomials.

Since the wave functions factorize, the operators $\XR{\R}$ will act only on
the $\R$-dependent part of it, having the same expressions as in (\ref{H-WtOp})
(changing there $\x$ to $\R$), and the operators $\XR{\x}$ will act only on the
$(\r,t)$-dependent part, with the expressions:
\bea
\XR{\kappa_1}\equiv\n\cdot\XR{\r} \Omega(\kappa_2,t)&=& \left[-\sin\wc
t\parcial{\kappa_2} +
     \idhbar m\wc\kappa_2\cos\wc t\right]\Omega(\kappa_2,t) \nn\\
\XR{\kappa_2}\equiv\n\cdot\J\cdot\XR{\r} \Omega(\kappa_2,t)&=& \left[\cos\wc
t\parcial{\kappa_2}
+     \idhbar m\wc\kappa_2\sin\wc t\right]\Omega(\kappa_2,t)
\eea

\ni once the (irrelevant) phase factors have been factorized out.

Using the dual transformation to
the one taking  $(\x,\p)$ to $(\r,\R)$, we obtain the expression of the
operators $\Xv$ and $\Pv$ in terms of $\XR{\r}$ and $\XR{\R}$:
\bea
\idhbar \Pv \equiv \XR{\x} &=& \XR{\r} \nn \\
-\idhbar \Xv\equiv \XR{\p} &=&
\frac{1}{m\wc}\J\cdot\left(\XR{\r}-\XR{\R}\right)
\eea

\ni In addition, by $\Tv$ we denote the operator $-i\hbar
\XR{\R}=\Pv-m\wc\J\cdot\Xv$.
It can be easily deduced that $\Pv$ has the physical meaning of a linear
momentum (mass times velocity), which we shall simply call momentum,
while  $\Tv$ is a momentum commuting with the Hamiltonian, generally called
magnetic translations, and this is associated with the coordinate $\R$ of the
centre of the circumferences. We can
still define another momentum in the theory, the canonical momentum, as
$\Piv\equiv -\frac{i\hbar}{2}\left(\XR{\r}+\XR{\R}\right)$, which has
the particularity that its components mutually commute, and,
as can be easily checked, is a proper translation generator: it is written
(for $t=0$) as $\vec{\nabla}_{\x}$ when acting on $\Phi(x,t)$ in
(\ref{BWF}) at $t=0$. Its explicit expression and that of $\Tv$ on $\Phi(x,t)$
are:
\bea
\Piv\Psi & = & \z \e^{-\frac{i}{2\hbar} \p\cdot\x}
    \left[-\frac{i\hbar}{2}\left(\M+\I\right) \cdot\vec{\nabla}
       + \frac{m\wc}{4}\left(\N-\J\right)\cdot\x \right]\Phi(\x,t) \nn\\
\Tv\Psi & = & \z \e^{-\frac{i}{2\hbar} \p\cdot\x} \left[-i\hbar\vec{\nabla}
       - \frac{m\wc}{2}\J\cdot\x \right]\Phi(\x,t)
\eea

The r\^ole of the different momenta can be clarified by introducing the
vector potential operator in the usual form,
$\Av\equiv-\frac{m\wc}{2}\J\cdot\Xv
= \frac{i\hbar}{2}\left(\XR{\r}-\XR{\R}\right)$. Then, the
canonical momentum is rewritten $\Piv=\Pv+\Av$, and $\Tv=\Pv+2\Av=\Piv+\Av$.
Then it is easy to verify that $\E=\frac{1}{2m}\Pv^2=
\frac{1}{2m}\left(\Piv-\Av\right)^2= \frac{1}{2m}\left(\Tv-2\Av\right)^2$.

\subsection{Charged particle in a homogeneous magnetic field in the
plane with periodic boundary conditions}

Before imposing the periodic boundary conditions which  define the torus,
as in Sec. 2.2, we
must determine how these boundary conditions affect each of the
coordinates. Clearly, $\x$ will be affected by the boundary
conditions, but it is not clear what happens to $\p$. Let us return to
$\r$ and $\R$ coordinates, where $\R$ is the (absolute) position of
the centre of the circumference (the classical trajectory) and $\r$ is the
(relative) position
of the particle with respect to the centre of the circumference, i.e.
$\r=\x-\R$. Therefore, $\R$ will be subject to periodic boundary
conditions (the same that $\x$) while $\r$ will not, being a
relative coordinate (since the classical energy $H$ is a function of
$\ro{}^2=\r{\:}^2$, periodic boundary conditions for $\r$ would imply an
upper bound to the energy spectrum, and even more, a periodic energy
spectrum). This makes $\r$ and $\R$ coordinates more
appropriate to describe the system with periodic boundary conditions.
Now we are ready to apply  the results of Sec. 2.2, having reduced the problem,
roughly speaking, to the study of an harmonic oscillator times a
Heisenberg-Weyl group on the torus, the latter being parameterized by $\R$.

Regarding the H-W subgroup, we can apply the results of Sec. 2.2.
We also consider the two cases {\it i)} and {\it ii)}, corresponding to $T$
being a trivial or non-trivial principal fibre bundle, respectively.

Let us consider first the case {\it i)} (Sec. 2.2.1), which is the more
conventional one. The actual condition to be satisfied is
\be
\frac{m\wc L_1L_2}{2\pi\hbar}=n\in Z \label{BQuant}
\ee

\ni which implies, as already anticipated, a quantization
of the magnetic flux through the torus surface, in the same manner as in
the Dirac monopole case. If this flux were produced by a monopole
charge, the quantization of the magnetic charge would follow. This kind of
quantization condition guarantees, for instance, that the Wilson loop
variables in gauge theories are single-valued \cite{Manton}.

The wave functions turn out to be (\ref{BWFr-R}), where $\Phi(y_2)$ is subject
to exactly the same restrictions as in Sec. 2.2.1, thus leading
to the expression (for $\n=(1,0)$):
\be
\Phi^{\av}(R_2)=\sum_{k=0}^{n-1}a_k \Lambda^{\av}_k(R_2)
\ee

\ni where $\av$ is defined as before. The wave function is therefore peaked at
$R_2=\a_2+\frac{k}{n}L_2,\, k\in Z$
($R_1=\a_2+\frac{k}{n}L_1,\, k\in Z$ for $\n=(0,1)$).

The subgroup $\GH$ of good
transformations (the ones that preserve the structure of the wave
function)  is the subgroup of \Gt with the parameters $\R$
restricted to be $\R=\frac{1}{n}\L_{\k}$. The quantum operators $\E$ and
$\XR{\r}$ are
good operators (since the harmonic oscillator part is not subject to
constraints),
while the operator $\XR{\R}$ is a bad operator.
If we analyse these results in
terms of the operators $\Xv,\, \Pv,\, \Tv\,$ and $\Piv$ by means of the
expressions given in Sec. 4.1, we conclude that the operator $\Pv$ is a
good operator, while $\Xv,\, \Tv\,$ and $\Piv$ are  not. Consequently, the
momentum
(or velocity) of the particle is a measurable quantity, but the position, the
canonical momentum and the magnetic translations are not observables. The
vector potential operator is of course  also a bad operator.
For all the bad operators, their finite expressions (counterparts of
$\hat{\eta}$ of Sec 2.1.1) can be nevertheless constructed, since all these
expressions are good operators.

For the case $ii)$, only the fractional case $a)$ is physically meaningful.
Now the wave function is defined on a torus $r$ times greater in one
direction \cite{Niu}, or, what is the same, it is a vector-valued (with $r$
components)
function; or, in FQHE terminology, the centre of mass function corresponding
to the vacuum is degenerate, $r$ being the degeneration.
In this case the Hall conductivity is associated with the quotient
$\frac{n}{r}$
of the Chern class of the associated determinant bundle by the rank of the
vector bundle \cite{determinante}.
This result lends
support to the idea that Fractional Quantum Hall Effect is always associated
with multiple-valued wave functions, i.e. degenerate vacua.

\section*{Acknowledgement}

We wish to thank Mark Gotay for valuable discussions.

\vfil\eject


\begin{thebibliography}{99}


\bibitem[{\bf A-A}]{GAQ} Aldaya, V., Azc\'arraga, J.A.: J. Math. Phys. {\bf
23},
              1297 (1982)

\bibitem[{\bf A-N-R}]{Alfonso}  Aldaya, V., Navarro-Salas, J., Ramirez, A.:
         Comm. Math. Phys. {\bf 121}, 541 (1989)

\bibitem[{\bf A-N-B-L}]{Schrodinger} Aldaya, V., Navarro-Salas, J., Bisquert,
J., Loll, R.:
                      J. Math. Phys. {\bf 33}, 3087 (1992)

\bibitem[{\bf A-B-G-N}]{Position} Aldaya, V., Bisquert, J., Guerrero, J.,
Navarro-Salas, J.:
                   J. Phys. {\bf A26}, 5375 (1993)


\bibitem[{\bf A}]{Ashtekar} Ashtekar, A.: {\ Mathematical problems of
non-perturbative
                  Quantum General Relativity}, Lectures delivered at the
                   1992 Les Houches summer school on Gravitation and Cosmology
                   (1992)

\bibitem[{\bf A-E-P}]{Asorey} Asorey, M., Esteve, J.G., Pacheco, A.F.:
                Phys. Rev. {\bf D27}, 1852 (1983)

\bibitem[{\bf C-D-L}]{Cohen-Tannoudji} Cohen-Tannoudji, C., Diu, B., Lalo\"e,
F.: {\it M\'ecanique quantique},
                          Tome I, Hermann (1977)

\bibitem[{\bf E}]{Esteve} Esteve, J.G.: Phys. Rev. {\bf D34}, 674 (1986)


\bibitem[{\bf F-F-Z}]{Fairlie} Fairlie, D.B., Fletcher, P., Zachos, C.K.:
                 J. Math. Phys. {\bf 31}, 1088 (1990)


\bibitem[{\bf F}]{Floratos} Floratos, E.G.: Phys. Lett. {\bf B 228}, 335 (1989)


\bibitem[{\bf G-G-H}]{Gotay} Gotay, M.J., Grundling, H., Hurst, C.A.: {\it A
Groenewold-Van
                Hove Theorem for $S^2$} (dg-ga/9502008)
(to appear in Transactions of the AMS) (1995)

\bibitem[{\bf G}]{GotayToro} Gotay, M.J., {\it On a Full Quantization of the
          Torus}, dg-ga/9507005 (1995)

\bibitem[{\bf I-L}]{Isham} Isham, C.J., Linden, N.: Class. Quant. Grav. {\bf
5},
                71 (1988)

\bibitem[{\bf K-D-P}]{Hall1}   Klitzing, K. v., Dorda, G., Pepper, M.:
                Phys. Rev. Lett. {\bf 45}, 494 (1980);

\bibitem[{\bf L-L}]{Landau}  Landau, L.D., Lifshitz, E.M.: {\it Quantum
Mechanics,
                       Nonrelativistic Theory}, Pergamon Press, Oxford
                       (1965)

\bibitem[{\bf L}]{Landsman} Landsman, N.P.: Let. Math. Phys. {\bf 20}, 11
(1990)


\bibitem[{\bf La}]{Hall2}   Laughlin, R.B.: Phys. Rev. Lett. {\bf 54}, 1395
(1983);


\bibitem[{\bf M}]{Manton}  Manton, N.S.: Ann. of Phys. {\bf 159}, 220 (1985)


\bibitem[{\bf N-T-W}]{Niu}  Niu, Q., Thouless, D.J., Wu, Y.S.: Phys. Rev. {\bf
B31}, 3372
               (1985)


\bibitem[{\bf O-K}]{O-K}  Ohnuki, Y., Kitakado, S.: J. Math. Phys. {\bf 34},
2827 (1993)


\bibitem[{\bf S}]{Shen} Shen, X.: Int. J. Mod. Phys. {\bf A 7}, 6953 (1992)

\bibitem[{\bf T}]{Thouless} Thouless, D.J.: J. Math. Phys. {\bf 35}, 5362
(1994)


\bibitem[{\bf V}]{determinante} Varnhagen, R.: {\it Topology and Fractional
Quantum Hall
                     Effect}, preprint BONN-Th-94-22, hep-th/9411040 (1994).


\bibitem[{\bf W}]{Weyl} Weyl, H.: {\it The Theory of Groups and Quantum
Mechanics},
               Dover, New York (1931)


\bibitem[{\bf Wo}]{Woodhouse} Woodhouse, N.: {\it Geometric Quantization},
Oxford
                    University Press, Oxford (1980)


\bibitem[{\bf W-Y}]{Wu-Yang}  Wu, T.T., Yang, C.N.: Phys. Rev. {\bf D 12}, 3845
(1975)

\bibitem[{\bf Z}]{Zak}  Zak, J.: Phys. Rev. {\bf 168}, 686 (1968)

\end{thebibliography}
\end{document}